\documentclass[%
superscriptaddress,
preprint,
 amsmath,amssymb,
prb,
]{revtex4-2}

\usepackage{dcolumn}
\usepackage[pdftex]{graphicx}
\usepackage{here}

\usepackage{bm}
\usepackage{comment}
\usepackage{xcolor}
    
\usepackage{amsmath, amssymb, amsfonts}

\usepackage[colorlinks=true,allcolors=blue]{hyperref}

\usepackage{siunitx} 
\NewCommandCopy\sqty\qty 
\NewDocumentCommand\Sqty{smm}{%
    \IfBooleanTF{#1}%
    {\sqty[quantity-product={}]{#2}{#3}}%
    {\sqty[quantity-product=~]{#2}{#3}}%
}
\newcommand{\mic}{\micro} 

\usepackage[italicdiff]{physics} 
\usepackage[version=4]{mhchem} 

\usepackage{cleveref}
\crefname{equation}{Eq.}{Eqs.} 
\crefname{figure}{Fig.}{Figs.}
\Crefname{figure}{Figure}{Figures}

\newcommand{\Ag}{A_{\mathrm{g}}}
\newcommand{\Au}{A_{u}}

\newcommand{\Eg}{E_{\mathrm{g}}}
\newcommand{\Eu}{E_{\mathrm{u}}}
\newcommand{\nEg}[1]{{}^{#1}\Eg}
\newcommand{\nEu}[1]{{}^{#1}\Eu}
\newcommand{\GMRL}{\Gamma^{\mathrm{RL}}}
\newcommand{\GMLR}{\Gamma^{\mathrm{LR}}}
\newcommand{\GMLL}{\Gamma^{\mathrm{LL}}}
\newcommand{\GMRR}{\Gamma^{\mathrm{RR}}}

\newcommand{\IRL}{I_{\mathrm{RL}}}
\newcommand{\ILR}{I_{\mathrm{LR}}}
\newcommand{\ILL}{I_{\mathrm{LL}}}
\newcommand{\IRR}{I_{\mathrm{RR}}}

\newcommand{\vbe}[1]{\vb{e}_{\mathrm{#1}}}

\newcommand{\Hopt}{H_{\mathrm{opt}}}
\newcommand{\Hep}{H_{\mathrm{ep}}}
\newcommand{\oT}{1T\textrm{-}}
\newcommand{\ddt}{d\textrm{-}d}

\newcommand{\transp}[1]{#1^\top} 
\NewDocumentCommand\qtxt{sm}{%
    \IfBooleanTF{#1}%
    {~\textrm{#2}}%
    {\quad\textrm{#2}}%
}
\newcommand{\pad}{\,}

\NewDocumentCommand{\subref}{mm}{\hyperref[#1]{\ref{#1}(#2)}}
\graphicspath{%
{./}%
}

\newcommand{\bk}{{\vb{k}}}
\newcommand{\bq}{{\vb{q}}}
\newcommand{\bA}{{\vb{A}}}
\newcommand{\bE}{{\vb{E}}}
\newcommand{\bJ}{{\vb{J}}}
\newcommand{\bc}{{\vb{c}}}
\newcommand{\PH}{{Q}}
\usepackage{braket}
\newcommand{\groa}{g_{\mathrm{ROA}}}

\usepackage[normalem]{ulem}
\DeclareRobustCommand{\eraser}{\bgroup\markoverwith{\textcolor{red}{\rule[.5ex]{2pt}{0.4pt}}}\ULon}
\DeclareRobustCommand{\eraseb}{\bgroup\markoverwith{\textcolor{blue}{\rule[.5ex]{2pt}{0.4pt}}}\ULon}
\newcommand{\addr}[1]{\textcolor{red}{#1}}
\newcommand{\addb}[1]{\textcolor{blue}{#1}}

\renewcommand{\eraser}[1]{}
\renewcommand{\eraseb}[1]{}
\renewcommand{\addb}[1]{#1}
\renewcommand{\addr}[1]{#1}

\begin{document}


\title{Raman Optical Activity Induced by Ferroaxial Order in \ce{NiTiO3}}

\author{Gakuto~Kusuno}
\affiliation{%
Department of Physics, Institute of Science Tokyo, Tokyo 152-8551, Japan%
}%
\author{Takeshi~Hayashida}
\affiliation{%
Department of Applied Physics, The University of Tokyo, Tokyo 113-8656, Japan%
}%
\affiliation{%
FELIX  Laboratory,  Radboud  University,  Toernooiveld  7,  6525  ED  Nijmegen,  The  Netherlands%
}%
\author{Takayuki~Nagai}
\affiliation{%
Department of Applied Physics, The University of Tokyo, Tokyo 113-8656, Japan%
}%
\author{Hikaru~Watanabe}
\affiliation{%
Department of Applied Physics, Hokkaido University, Sapporo 060-8628, Japan%
}
\affiliation{%
Department of Physics, The University of Tokyo, Tokyo 113-0033, Japan%
}%
\author{Rikuto~Oiwa}
\affiliation{%
Department of Physics, Hokkaido University, Sapporo 060-0810, Japan%
}
\affiliation{Center for Emergent Matter Science, RIKEN, {Wako}, Saitama 351-0198, Japan}
\author{Tsuyoshi~Kimura}
\affiliation{%
Department of Applied Physics, The University of Tokyo, Tokyo 113-8656, Japan%
}%
\author{Takuya~Satoh}
\affiliation{%
Department of Physics, Institute of Science Tokyo, Tokyo 152-8551, Japan%
}%
\affiliation{%
Quantum Research Center for Chirality, Institute for Molecular Science, Aichi 444-8585, Japan%
}%

\date{\today}

\begin{abstract}
  Raman optical activity (ROA), the dependence of Raman intensity on the circular polarization of incident and scattered light, has traditionally been \eraseb{\addb{associated with}}\addb{observed in} chiral molecules and magnetic materials\addb{,} \addb{where inversion or time-reversal symmetry is broken}. 
  \eraseb{\addb{In this study,}}\addb{Here} we demonstrate that ROA can also arise in \eraseb{\addb{ferroaxial materials that possess spatial inversion and time-reversal symmetries}}\addb{a centrosymmetric and non-magnetic ferroaxial crystal}. 
  Using circularly polarized Raman spectroscopy on single-crystalline $\ce{NiTiO3}$, we observed a pronounced ROA signal in the cross-circular polarization configuration\addb{s}, which correlates with the ferroaxial domain structure. 
  Our symmetry analysis\addb{,} \addb{first-principles calculations of phonons,} and tight-binding model calculations reveal that the natural ROA\eraseb{\addb{ (NROA)}} originates from the ferroaxial order and persists even within the electric dipole approximation. 
  These results establish ROA as a powerful probe of ferroaxial order in centrosymmetric systems.
\end{abstract}

\maketitle


Chirality is the property of objects that cannot be superimposed on their mirror images by rotation, as exemplified by \eraseb{\addb{right}}\addb{left} and \eraseb{\addb{left}}\addb{right} hands. 
It is widespread in nature and occurs in crystals with helical structures. 
When light enters such chiral materials, it gives rise to optical rotation and circular dichroism, phenomena collectively known as natural electronic optical activity (NEOA)~\cite{barron2004z}. 
The \eraseb{\addb{angle}}\addb{sense} of rotation of linearly polarized light and the differential absorption of \eraseb{\addb{right-}}\addb{left-} and \eraseb{\addb{left-}}\addb{right-}circularly polarized light vary with the handedness of the chiral material. 
Therefore, NEOA serves as a valuable technique for probing chirality.

A similar phenomenon occurs in Raman scattering, known as Raman optical activity (ROA), which involves differences in the Raman intensities depending on the circular polarization state of the incident and scattered light. 
ROA has been reported in systems containing chiral molecules and in those with magnetic ordering. 
ROA arising from magnetic effects is referred to as magnetic ROA, while that occurring in non-magnetic materials is called natural ROA (NROA)~\cite{barron2004z}.
The NROA of chiral molecules is attributed to the interference between scattered light resulting from electric dipole-electric dipole transitions and that from electric dipole-magnetic dipole transitions~\cite{barron1971}. 
For NROA to occur, \eraser{\addr{the material must lack spatial inversion symmetry}}\addr{the materials have traditionally been thought to lack inversion symmetry}.

However, recent experimental studies suggest that NROA can also be observed in \eraseb{\addb{non-magnetic}}\addb{centrosymmetric,} \eraseb{\addb{and }}\eraseb{\addb{centrosymmetric}}\addb{non-magnetic} materials. 
In the layered compound $\oT$\ce{TaS2}, intensity differences between \eraseb{\addb{right-}}\addb{left-} and \eraseb{\addb{left-}}\addb{right-}circularly polarized Raman spectra have been observed in the chiral charge density wave (CDW) phase~\cite{lacinska2022,yang2022b,zhao2023b,liu2023}. 
Despite its name, the chiral CDW phase of $\oT$\ce{TaS2} is not strictly chiral, as its point group, $\bar{3}$, is achiral and retains inversion symmetry. 
This state is more accurately described as exhibiting ferroaxial (\eraseb{\addb{planar chiral or }}ferro-rotational) order~\cite{liu2023}.
Previous studies have attributed the origin of ROA in this material to chiral CDW domains formed through complex Raman tensors owing to optical absorption~\cite{yang2022b}. 
Similar circular intensity differences in Raman scattering have also been reported in \eraseb{\addb{chiral and polar compounds}}\addb{compounds that are both chiral and polar,} such as \ce{NiCo2TeO6}, which belongs to the ferroaxial point group $3$~\cite{martinez2025}. 
However, the microscopic mechanism by which NROA arises from ferroaxial order remains unknown.

In this study, we report the observation of a \eraseb{\addb{giant}}\addb{remarkably large} ROA in a three-dimensional centrosymmetric bulk crystal of \ce{NiTiO3}, a material known to exhibit ferroaxial order. 
The intensity differences observed in both the Stokes and anti-Stokes spectra are consistent with the symmetry arguments of NROA.
Furthermore, our \eraseb{\addb{model }}calculations show that the observed NROA originates directly from the ferroaxial order.

Ferroaxial order was proposed as a type of ferroic order that is invariant under both spatial inversion and time-reversal \eraseb{\addb{symmetries}}\addb{operations}~\cite{johnson2011}. 
It arises from the breaking mirror symmetry, with the mirror plane(s) obtained parallel to the principal axis. 
The order parameter is the axial vector $\vb{A}\propto\sum_i\vb{r}_i\cp\vb{p}_i$, where $\vb{r}_i$ and $\vb{p}_i$ represent the position vector and electric dipole moment of the $i$th atom, respectively~\cite{cheong2018}. 
Such axial vectors are allowed in only 13 of the 32 crystal point groups~\cite{hlinka2016,hayami2018}.
Since chirality-related phenomena, such as NEOA, require the breaking of inversion symmetry, some ferroaxial materials with centrosymmetric structures do not exhibit NEOA. 
Well-known ferroaxial systems that retain inversion symmetry include \ce{RbFe(MoO4)2}~\cite{jin2020}, \ce{K2Zr(PO4)2}~\cite{yamagishi2023,bhowal2024a}, \ce{NiTiO3}~\cite{hayashida2020,yokota2022,guo2023a,hayashida2023}, and \ce{MnTiO3}~\cite{sekine2024,zhang2025c}.

In \ce{NiTiO3}, the high-temperature phase above the transition temperature $T_{\mathrm{C}}=\Sqty{1560}{K}$ adopts a corundum-type structure, where \ce{Ni^{2+}} and \ce{Ti^{4+}} ions are randomly distributed at the cation sites [Fig.~\subref{fig:structure}{a}]. 
The space group is $R\bar{3}c$ (point group $\bar{3}m$), and the structure lacks ferroaxial order due to the presence of a glide (mirror) plane along the principal axis. 
As the temperature decreases, a structural phase transition occurs at $T_{\mathrm{C}}$~\cite{lerch1992,hayashida2020}. 
Below $T_{\mathrm{C}}$, the \ce{Ni^{2+}} and \ce{Ti^{4+}} ions become ordered. 
The space group changes to $R\bar{3}$ (point group $\bar{3}$), breaking the mirror symmetry. 
Consequently, a net axial vector emerges. 
The $A_+$ and $A_-$ domain states shown in Figs.~\subref{fig:structure}{b} and \subref{fig:structure}{c}, respectively, correspond to different axial vector orientations and are related by a $c$-glide operation or by flipping the crystal along the $c$-axis.
\begin{figure}
  \includegraphics[width=1.0\hsize]{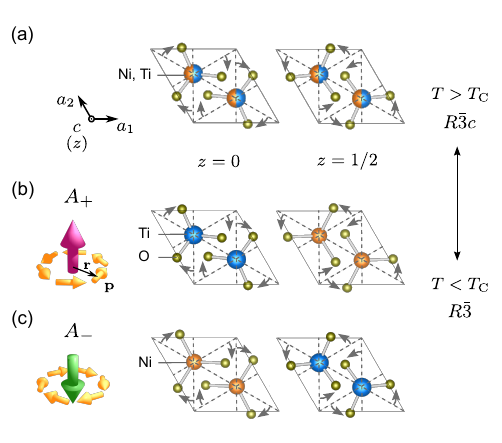}
  \caption{\label{fig:structure}
  (a) Crystal structure of \ce{NiTiO3} above the structural phase transition temperature ($T_{\mathrm{C}} = \Sqty{1560}{K}$). 
  Ti and Ni cations on $z = 0$ and $z = 1/2$ layers and oxygen ions bonded to them are depicted.
  (b), (c) Crystal structures viewed along the $c$-axis for the $A_+$ and $A_-$ ferroaxial domains, respectively.
  The \addb{grey} arrows denote the direction of rotational displacements of oxygen ions from the (110)-type planes (dotted lines).
  The ferroaxial order\addb{, }\addb{indicated by the purple or green arrows,} is ch\addb{a}racterized as $\vb{A} \propto \sum_i \vb{r}_i \times \vb{p}_i$,
  where $\vb{r}_i$ and $\vb{p}_i$ are the position \addb{vector} and \addb{the} electric dipole moment \addb{(yellow arrows)} at site $i$\addb{, }\addb{respectively }\addr{(see Supplemental Material, Sec.~S1 for details)}. 
  }
\end{figure}

We prepared both single- and multidomain crystals of \ce{NiTiO3}. 
A single-domain crystal grown using the flux method~\cite{garton1972} was \Sqty{60}{\mic m} thick and was confirmed to be a single-domain by electrogyration measurements~\cite{hayashida2020,hayashida2021}. 
A multidomain crystal was obtained by heating a single-domain crystal to \Sqty{1620}{K} and then slowly cooling it, resulting in a mixture of ferroaxial domains.
This domain structure was also confirmed via electrogyration, as reported in Ref.~\cite{hayashida2021}.

\eraseb{\addb{Circularly polarized Raman spectroscopy was conducted}}\addb{We performed circularly polarized Raman spectroscopy} in a backscattering geometry with excitation along the $c$-axis using \Sqty{785}{nm}\eraser{\addr{ and}}\addr{,} \Sqty{532}{nm}\addr{, and \Sqty{633}{nm}} lasers. 
\addb{The detailed setup is shown in Supplemental Material~\cite{supple}, Sec.~S2.}
In the wavelength range around \Sqty{785}{nm}, a strong absorption occurs due to $\ddt$ transitions of the \ce{Ni^2+} ions, 
whereas the absorption \eraseb{\addb{is relatively weak }}at \Sqty{532}{nm} \addb{is much weaker and originates from the tail of a charge-transfer transition band}. 
\addr{At \Sqty{633}{nm}, the absorption is minimal~\cite{li2016}.} 
The scattered light was collected through an objective lens, filtered to remove Rayleigh scattering using \addb{edge/}notch filters, and detected with a CCD-equipped spectrometer. 
All measurements were performed at \Sqty{295}{K}\addb{, }\addb{which is far above the antiferromagnetic N\'eel temperature, $T_{\mathrm{N}}=\Sqty{22.5}{K}$~\cite{dey2020b}}.
\eraseb{\addb{Right (R)}}\addb{Left (L)} and \eraseb{\addb{left (L)}}\addb{right (R)} circular polarizations were defined based on the rotation of the electric field on the sample plane, independent of the propagation direction. 
Four polarization configurations for the incident and scattered light were considered: \eraseb{\addb{RL}}\addb{LR} and \eraseb{\addb{LR}}\addb{RL} (cross-circular) and LL and RR (parallel-circular), as shown in Figs.~\subref{fig:mode_assign}{a} and \subref{fig:mode_assign}{b}, respectively. 
According to the polarization selection rules for Raman-active modes $\Gamma = 5\Ag \oplus 5\Eg$, $\Eg$-mode phonons are active in cross-circular \addb{polarization} configurations, whereas $\Ag$ modes are active in parallel-circular \addb{polarization} configurations \addb{(see Supplemental Material~\cite{supple}, Sec.~S3 for details)}.
\begin{figure}
  \includegraphics[width=1.0\hsize]{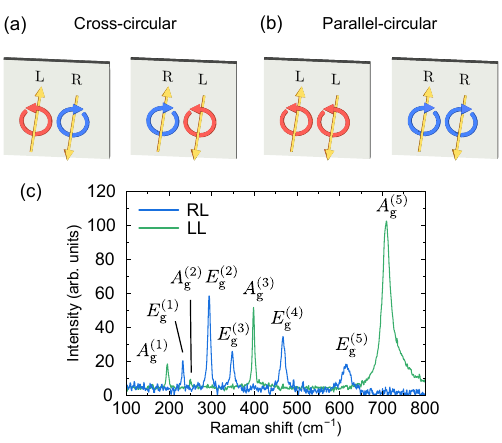}
  \caption{\label{fig:mode_assign}
  (a), (b) Experimental setups for circularly polarized Raman spectroscopy in the cross-circular and parallel-circular polarization configurations, respectively. 
  (c) Raman spectra of the single-domain \ce{NiTiO3} crystal, 
  measured at \Sqty{295}{K} using \Sqty{785}{nm} excitation in RL (R-incident, L-scattered) and LL (L-incident, L-scattered)\eraseb{\addb{ polarization}} configurations, with phonon mode assignments.}
\end{figure}


Raman spectra obtained from the front surface of the single-domain crystal at an excitation wavelength of \Sqty{785}{nm} for the RL and LL configurations are shown in Fig.~\subref{fig:mode_assign}{c}.
Four $\Ag$ and five $\Eg$ modes were identified. 
\addb{The assignment of} \eraseb{\addb{T}}\addb{t}hese modes agree\addb{s} well with those reported in previous studies on \ce{NiTiO3} powders~\cite{fujioka2016, qi2022}.
Although the $\Ag^{(4)}$ mode has been reported to exist near \Sqty{480}{cm^{-1}} and was confirmed by our first-principles calculations (\addb{see} Supplemental Material~\cite{supple}, Sec.~S4 \addb{for details}), it was not experimentally observed due to its weak intensity.

Next, we conducted measurements in the cross-circular polarization \addb{(LR and RL)} configuration\addb{s}.
The Raman spectra from the front surface of the single-domain crystal at an excitation wavelength of \Sqty{785}{\nm} are shown in Figs.~\subref{fig:785spectrum}{a} and \subref{fig:785spectrum}{b}, respectively.
\addb{All} \eraseb{\addb{F}}\addb{f}ive phonon peaks corresponding to $\Eg$ modes are observed in both the Stokes and anti-Stokes spectra.
These peaks exhibit clear intensity differences between the \eraseb{\addb{RL}}\addb{LR} and \eraseb{\addb{LR}}\addb{RL} configurations. 
This result \eraseb{\addb{indicates}}\addb{demonstrates} that ROA occurs even in a centrosymmetric, achiral crystal\addb{,} \ce{NiTiO3}.
To quantify the ROA, we defined the normalized intensity difference between the cross-circular \addb{polarization} configurations as
\begin{align}
  \groa=\frac{\ILR-\IRL}{(\ILR+\IRL)/2},
\end{align}
where $\ILR$ and $\IRL$ are the Raman intensities of the LR and RL configurations, respectively.
The value of $\groa$ varies across the phonon modes, with the $\Eg^{(1)}$ mode exhibiting the largest intensity difference, reaching $\groa \simeq 1.0$.
This value is several orders of magnitude larger than the typical NROA of chiral molecules, where $\groa \sim 10^{-3}$~\cite{barron1971} 
(see Supplemental Material~\cite{supple}, \addb{Table~S4 in} Sec.~S\eraseb{\addb{3}}\addb{5}, for the values of $\groa$ for the other peaks).
\addb{The identical sign of $\groa$ in the Stokes and anti-Stokes spectra indicates that the observed ROA in \ce{NiTiO3} is non-magnetic in origin, in contrast to magnetic ROA where the sign reverses~\cite{barron1976,barron1982,cenker2021}.}
\begin{figure*}
  \includegraphics[width=1.0\hsize]{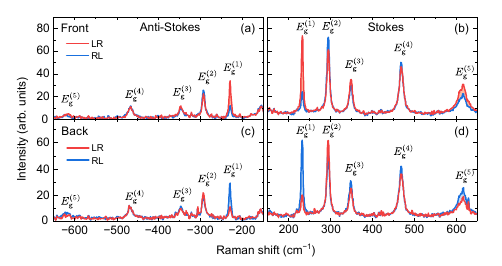}
  \caption{\label{fig:785spectrum}
  Raman spectra of the front (a, b) and back (c, d) surfaces of the single-domain \ce{NiTiO3} crystal, 
  measured at \Sqty{295}{K} using \Sqty{785}{nm} excitation in \eraseb{\addb{RL}}\addb{LR} and \eraseb{\addb{LR}}\addb{RL} \eraseb{\addb{polarization }}configurations. 
  (a), (c) show anti-Stokes spectra, while (b), (d) show Stokes spectra.}
\end{figure*}

Figures~\subref{fig:785spectrum}{c} and \subref{fig:785spectrum}{d} show Raman spectra measured on the back surface of the same single-domain crystal.
We again observed intensity differences between the \eraseb{\addb{RL}}\addb{LR} and \eraseb{\addb{LR}}\addb{RL} configurations; however, the sign of $\groa$ was reversed compared to the front side.
This indicates that the observed NROA in \ce{NiTiO3} is direction\eraseb{\addb{al}}\addb{-dependent}---i.e., it \addr{depends on the orientation (sign) of the ferroaxial vector}\eraser{\addr{ behaves as a vector quantity}}.
Such directionality does not originate from the bulk properties of chiral materials.
Instead, it reflects the domain-dependent orientation of the ferroaxial order, which differs between the front and back of the crystal.
\addr{The differential Raman spectra of cross-circular polarization configurations on the front and back surfaces are shown in Fig.~S6 of Supplemental Material~\cite{supple}, Sec.~S6.}

In contrast to the $\Eg$ modes, no noticeable intensity differences were observed for the $\Ag$ modes in the parallel-circular polarization \addb{(LL and RR)} configurations (\addb{Fig.~S7 in} Supplemental Material~\cite{supple}, Sec.~S\eraseb{\addb{1}}\addb{7}). 
\eraseb{\addb{This suggests that NROA is selectively enhanced in certain phonon modes due to coupling with the electronic structure associated with the ferroaxial order.}}
Moreover, the $\Eg$ modes excited at a wavelength of \Sqty{532}{nm} exhibited only \eraseb{\addb{minor}}\addb{weak} intensity differences \addr{and no discernable intensity differences at \Sqty{633}{nm}} (see Supplemental Material~\cite{supple}, Sec.~S\eraseb{\addb{2}}\addb{5}).
\addb{These results suggest that NROA is selectively enhanced in certain phonon modes due to coupling with the electronic structure associated with the ferroaxial order.}

\addr{The selective enhancement of ROA at \Sqty{785}{nm} can be understood as a cooperative effect of (i) symmetry and (ii) resonance.}
\addr{From (i) the symmetry perspective, the $\Eg$ phonons decompose into two components, $\nEg{1}$ and $\nEg{2}$, corresponding to counterclockwise and clockwise rotational motions with crystal (pseudo)-angular momenta of $+1$ and $-1$, respectively. 
Under circularly polarized excitation, the LR and RL configurations transfer angular momenta of $-2$ and $+2$ to the material, thereby selectively exciting the $\nEg{1}$ and $\nEg{2}$ phonons in accordance with angular momentum conservation and the Umklapp process in the three-fold symmetry~\cite{ishito2023a}.}

The Raman intensity in the electric dipole approximation can be expressed as
$I\propto \qty|\vbe{s}^\dagger R\vbe{i}|^2$,
where $\vbe{i}$ and $\vbe{s}$ are the polarization vectors of the incident and scattered light, respectively, and $R$ is the Raman tensor.
The Raman tensor for the $\Eg$ phonons are given by
\begin{align}
  R=R_{1}\mqty(1 & i \\
  i & -1 )
  +R_{2}\mqty(1 & -i \\
  -i & -1 ),
  \label{eq:RLLR_tensor}
\end{align}
where $R_{1}$ and $R_{2}$ are proportional coefficients of the Raman tensor for phonon modes belonging to \eraseb{\addb{irreducible representations }}${}^1\Eg$ and ${}^2\Eg$, respectively (see Supplemental Material~\cite{supple}, Sec.~S\eraseb{\addb{5}}\addb{3} \addb{for derivations}).
Since the Raman intensities in the LR and RL configurations are proportional to $\qty|R_{1}|^2$ and $\qty|R_{2}|^2$, respectively, 
a nonzero ROA arises when $|R_{1}| \neq |R_{2}|$.
\addr{The static ferroaxial distortion provides a preferred sense of rotation in the lattice, rendering the two rotational components, $\nEg{1}$ and $\nEg{2}$, inequivalent and leading to asymmetric Raman tensor elements ($\qty|R_{1}|\neq\qty|R_{2}|$), which in turn give rise to a finite ROA signal.}


\addr{For (ii) resonance, at \Sqty{785}{nm} the incident light is resonant with the \ce{Ni^{2+}} $\ddt$ transition. This transition is electric-dipole allowed because the local inversion symmetry at the \ce{Ni} site is already broken. 
The ferroaxial rotational distortion further lowers the local symmetry from $3m$ to $3$, which modifies the symmetry of the lattice displacements coupled to the $\ddt$ transition. Under this resonance, the intermediate electronic states involved in Raman scattering are highly sensitive to local lattice distortions. 
Our first-principles calculations show that the $\Eg^{(1)}$ mode involves the largest displacement of \ce{Ni^{2+}} ions, leading to the strongest coupling to the resonant transition and accounting for the largest observed ROA at \Sqty{785}{nm}.}

\addb{We verified the condition $|R_{1}| \neq |R_{2}|$ in ferroaxial systems using symmetry-based model calculations. 
Specifically, we consider a tight-binding Hamiltonian for spinless $p$ orbitals on a trigonal lattice with symmetry-adapted bases~\cite{kusunose2023}} 
\addr{and demonstrate that ferroaxial rotation generically leads to different scattering amplitudes in the two cross-circular polarization configurations, independent of microscopic band details.}
%
Regarding the Raman tensors $R(\omega, \delta \omega)$ with their explicit dependence on the frequency of the incident light $\omega$ and the (anti-)Stokes shift $\delta \omega$, we numerically computed the $\omega$ dependence of the coefficients $R_{i}$ ($i=1,2$).
Following the established perturbation calculations within the electric dipole approximation, we obtained the normalized cross-circular Raman intensities, defined as
\begin{equation}
    \overline{I}_\text{LR} = \frac{|R_{1}|^2}{|R_{1}|^2 + |R_{2}|^2},~\overline{I}_\text{RL} = \frac{|R_{2}|^2}{|R_{1}|^2 + |R_{2}|^2}.
\end{equation}
In Fig.~\ref{fig:model_stokes}, we observe a significant deviation between the two intensities ($\overline{I}_\text{LR} \neq \overline{I}_\text{RL}$), indicating a cross-circular ROA even within the electric dipole approximation.
Analysis of the essential parameters~\cite{oiwa2022a} shows that the cross-circular ROA results from the coupling of the $E_g$ phonons to the orbital rotations driven by the ferroaxial order (see Supplemental Material~\cite{supple}, Sec.~S\eraseb{\addb{6}}\addb{8} \addb{for details}).      
ROA is strongly enhanced in the presence of resonant particle-hole excitations, as illustrated by the joint density of states (JDOS), supporting the experimental results obtained with different light sources (Fig.~\ref{fig:785spectrum} and Supplemental Material~\cite{supple}, Sec.~S\eraseb{\addb{2}}\addb{5}).
Notably, consistent with experimental observations, the signs of the cross-circular ROA ($\overline{I}_\text{LR} - \overline{I}_\text{RL}$) are the same for the Stokes ($\delta \omega <0$) and anti-Stokes ($\delta \omega >0$) responses for each excitation frequency $\omega$ (Fig.~\ref{fig:model_stokes} and inset).
\addr{Note that although the model calculation was conducted on a trigonal lattice,
our symmetry analysis can extend to ferroaxial systems with $n$-fold rotational symmetry ($n = 1,\pad 2,\pad 3,\pad 4,\pad 6$), provided that the Raman susceptibility differs between the two cross-circular polarization configurations.}
\begin{figure}
  \includegraphics[width=1.0\hsize]{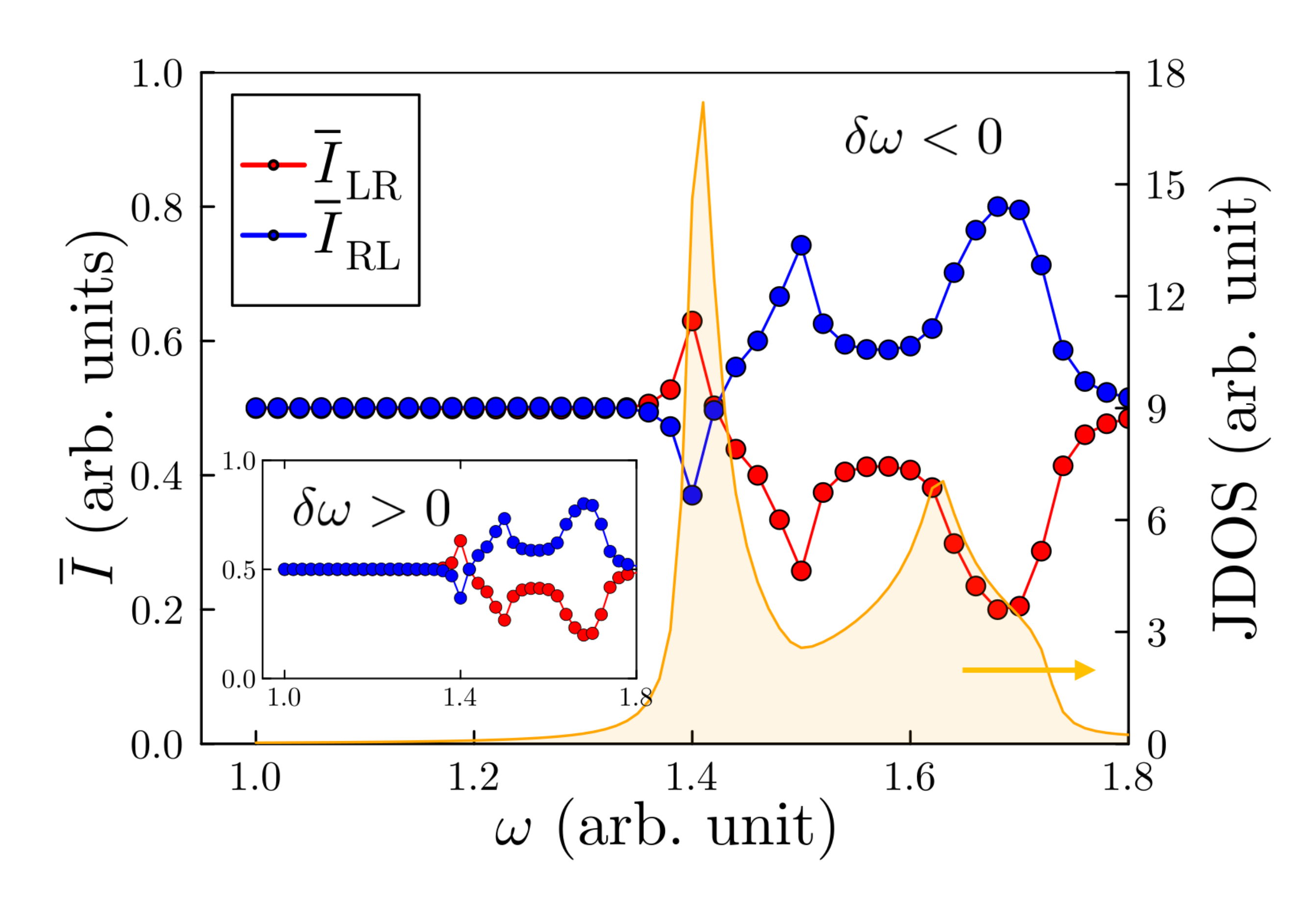}
  \caption{\label{fig:model_stokes}
  Excitation frequency ($\omega$) dependence of the normalized cross-circular Raman scattering intensities ($\overline{I}_\text{LR},\overline{I}_\text{RL}$) and the joint density of states (JDOS) obtained from model calculations.
  JDOS indicates the number of optically resonant states contributing to the Raman scattering.
  Main plots of $\overline{I}_\text{LR},\overline{I}_\text{RL}$ is for the Stokes process ($\delta \omega =-10^{-3}<0$), and the inset is for the anti-Stokes process ($\delta \omega =10^{-3}>0$).
  }
\end{figure}


To further confirm that the observed intensity differences reflect the ferroaxial order in \ce{NiTiO3}, we performed a spatial mapping of $\groa$ by scanning the laser spot across the sample.
Figures~\subref{fig:imaging}{a} and \subref{fig:imaging}{b} show spatial maps of $\groa$ near the $\Eg^{(1)}$ peak on the front and back surfaces of a single-domain crystal, respectively.
The sign of $\groa$ is reversed between the two surfaces, which is consistent with the change in the ferroaxial domain orientation.
In the regions outside the sample or near cracks, $\groa \sim 0$, indicating that the observed ROA was intrinsic and not due to instrumental offsets.
Figure~\subref{fig:imaging}{c} shows a similar mapping for a multidomain crystal, revealing a domain pattern with features on the order of \Sqty{100}{\mic m}.
This pattern closely matches the ferroaxial domain distribution obtained from electrogyration imaging [Fig.~\subref{fig:imaging}{d}] as reported in Ref.~\cite{hayashida2021}, which supports the conclusion that the ROA reflects the ferroaxial order.
\addr{The narrow region where $\groa$ approaches zero likely reflects either partial averaging of $A_{+}$ and $A_{-}$ domains within the optical spot or a suppression of the ferroaxial order at the domain wall; 
higher spatial resolution would be required to distinguish these scenarios (see Supplemental Material~\cite{supple}, Sec. S9).}
\begin{figure}
  \includegraphics[width=1.0\hsize]{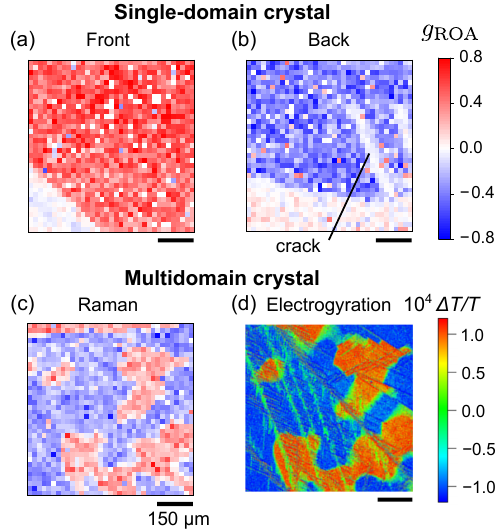}
  \caption{\label{fig:imaging}
  Imaging of ferroaxial domains in \ce{NiTiO3}. 
  (a), (b) Domain maps of the front and back surfaces of a single-domain crystal obtained by circularly polarized Raman spectroscopy. 
  (c) Domain map of a multidomain crystal. 
  The normalized intensity difference $\groa = 2(\ILR - \IRL)/(\ILR + \IRL)$ of the $\Eg^{(1)}$ peak is plotted. 
  (d) Domain image obtained by electrogyration, adapted from Ref.~\cite{hayashida2021} (image inverted from the original).}
\end{figure}

In summary, we demonstrated \eraseb{\addb{the}}\addb{a} \eraseb{\addb{giant}}\addb{remarkably large} ROA in the ferroaxial crystal \ce{NiTiO3} using circularly polarized Raman spectroscopy.
\addb{The intensity differences observed in the Stokes and anti-Stokes spectra, together with their dependence on the domain orientation, provide clear evidence that the ROA is induced by ferroaxial order.}
Our \eraseb{\addb{model}}\addb{calculations} show\eraseb{s} that this effect arises from \eraseb{\addb{the }}intrinsic electron\eraseb{\addb{ic}}--\eraseb{\addb{vibrational}}\addb{phonon} coupling within the electric dipole approximation, even in centrosymmetric, achiral, and non-magnetic materials.
These findings pave the way for the observation of ferroaxial domains using Raman spectroscopy and broaden the scope of ROA beyond chiral and magnetic systems.
Furthermore, the coupling between electrons and chiral phonons in \ce{NiTiO3} is expected to give rise to diverse phenomena, including phonon magnetic moments~\cite{juraschek2017,lujan2024}.

The data that support the findings of this study are available in the Supplemental Material.

We thank K. T. Yamada for technical assistance, and H. Kusunose, H. Yokota, and R. Arita for valuable discussions.
T.S. was supported in part by JSPS KAKENHI (Grants No. JP21H01032, No. JP22H01154, and No. JP26H02234), 
MEXT X-NICS (Grant No. JPJ011438), NINS OML Project (Grant No. OML012301), 
\eraseb{\addb{and }}JST CREST (Grant No. JPMJCR24R5)\addb{, and JST ERATO (Grant No. JPMJER2503)}.
T.K. was supported in part by JSPS KAKENHI (Grants No. JP25H00392 and No. JP25H01247).
H.W. was supported by JSPS KAKENHI (Grants No.~JP23K13058,~No. JP24K00581, and No. JP25H02115) and by RIKEN TRIP initiative (RIKEN Quantum, Advanced General Intelligence for Science Program, Many-body Electron Systems).
R.O. was supported by Special Postdoctoral Researcher Program at RIKEN.
\addb{G.K. was supported by JST SPRING (Grant No. JPMJSP2180) and the Science Tokyo Support Program for Doctoral Students, funded by the Universities for International Research Excellence.}
We used VESTA~\cite{momma2011} to visualize the crystal structures.


\bibliography{NiTiO3_paper.bib,NiTiO3_book.bib}

\clearpage

{

\title{Supplemental Material\addb{: }\\
Raman Optical Activity Induced by Ferroaxial Order in \ce{NiTiO3}}

\author{Gakuto~Kusuno}
\affiliation{%
Department of Physics, Institute of Science Tokyo, Tokyo 152-8551, Japan%
}%
\author{Takeshi~Hayashida}
\affiliation{%
Department of Applied Physics, The University of Tokyo, Tokyo 113-8656, Japan%
}%
\affiliation{%
FELIX  Laboratory,  Radboud  University,  Toernooiveld  7,  6525  ED  Nijmegen,  The  Netherlands%
}%
\author{Takayuki~Nagai}
\affiliation{%
Department of Applied Physics, The University of Tokyo, Tokyo 113-8656, Japan%
}%
\author{Hikaru~Watanabe}
\affiliation{%
Department of Applied Physics, Hokkaido University, Sapporo 060-8628, Japan%
}
\affiliation{%
Department of Physics, The University of Tokyo, Tokyo 113-0033, Japan%
}%
\author{Rikuto~Oiwa}
\affiliation{%
Department of Physics, Hokkaido University, Sapporo 060-0810, Japan%
}%
\affiliation{Center for Emergent Matter Science, RIKEN, {Wako}, Saitama 351-0198, Japan}
\author{Tsuyoshi~Kimura}
\affiliation{%
Department of Applied Physics, The University of Tokyo, Tokyo 113-8656, Japan%
}%
\author{Takuya~Satoh}
\affiliation{%
Department of Physics, Institute of Science Tokyo, Tokyo 152-8551, Japan%
}%
\affiliation{%
Quantum Research Center for Chirality, Institute for Molecular Science, Aichi 444-8585, Japan%
}%


\maketitle


\renewcommand{\theequation}{S\arabic{equation}}
\setcounter{figure}{0}
\renewcommand{\thefigure}{S\arabic{figure}}
\setcounter{table}{0}
\renewcommand{\thetable}{S\arabic{table}}
\setcounter{section}{0}
\renewcommand{\thesection}{S\arabic{section}}

\begin{center}
  \large\textbf{SUPPLEMENTAL MATERIAL}
\end{center}

\section{\NoCaseChange{\addr{Ferroaxial order in \ce{NiTiO3}}}}
\addr{In Figs.~1(b) and 1(c), we show a simplified crystal structure in each phase of \ce{NiTiO3} and depict only two \ce{TiO3} (\ce{NiO3}) triangular pyramids located at $z \approx 0$ for $A_{+}$ ($A_{-}$) domain and $z \approx 1/2$ for $A_{-}$ ($A_{+}$) domains. 
The $A_{+}$ and $A_{-}$ domains are related to each other by the operations whose symmetries are lost at the ferroaxial transition, such as the $c$-glide operation with the glide plane parallel to the (110) plane. 
Grey round arrows denote the direction of (virtual) displacements of oxygen ions from the (110)-type planes (dotted lines).}

\addr{The emergence of ferroaxial order can be understood in terms of virtual electric dipoles as follows.
In the average structure obtained by superposing the $A_{+}$ and $A_{-}$ domains, the oxygen ions lie on the (110)-type planes.
Deviations of the oxygen positions in the $A_{+}$ (or $A_{-}$) structure from these average positions give rise to virtual electric dipoles, which form a swirling pattern around the \ce{Ti} and \ce{Ni} ions.
}

\addr{In the $A_{+}$ ($A_{-}$) domain, the lengths of \ce{Ti}-\ce{O} bonds at $z \approx 0$ ($1/2$) and \ce{Ni}-\ce{O} bonds at $z \approx 1/2$ ($0$) are different, leading to a finite net axial vector $\vb{A}$ in a unit cell. 
By contrast, in the disordered nonferroaxial structure [Fig. 1(a)] with the same (\ce{Ti}, \ce{Ni})-\ce{O} bond lengths at $z \approx 0$ and $z \approx 1/2$, net $\vb{A}$ is cancelled out in a unit cell. 
This perspective reasonably explains the ferroaxial feature of \ce{NiTiO3}. 
}

\clearpage
\section{\NoCaseChange{\addr{Optical setup of Raman measurements}}}
\addr{Figure \ref{fig:setup} shows the experimental setup for circularly polarized Raman spectroscopy.
The incident laser light is first linearly polarized by a polarizer (P) and then directed to the sample through a beam splitter (BS) and an objective lens (OL). The backscattered light passes again through the beam splitter and is analyzed by an analyzer.
Circular polarization of both the incident and scattered light is controlled by inserting a quarter-wave plate (QWP) between the beam splitter and the objective lens.
Cross-circular polarization (LR and RL) configurations are obtained by setting the polarizer and analyzer in a parallel polarization configuration, and rotating the optic axis of the quarter-wave plate by \ang{45} with respect to the linear polarization direction.
Parallel-circular polarization (LL and RR) configurations are obtained by setting the polarizer and analyzer in a crossed polarization configuration, and again rotating the quarter-wave plate by \ang{45}.
The scattered light is filtered using an edge/notch filter (EF/NF), and then focused onto a pinhole (PH) by a lens (L) to improve depth resolution in a confocal configuration. Finally, the light is spectrally dispersed by a spectrometer and detected by a CCD sensor.}

\begin{figure}[H]
  \centering
  \includegraphics[width=0.9\hsize]{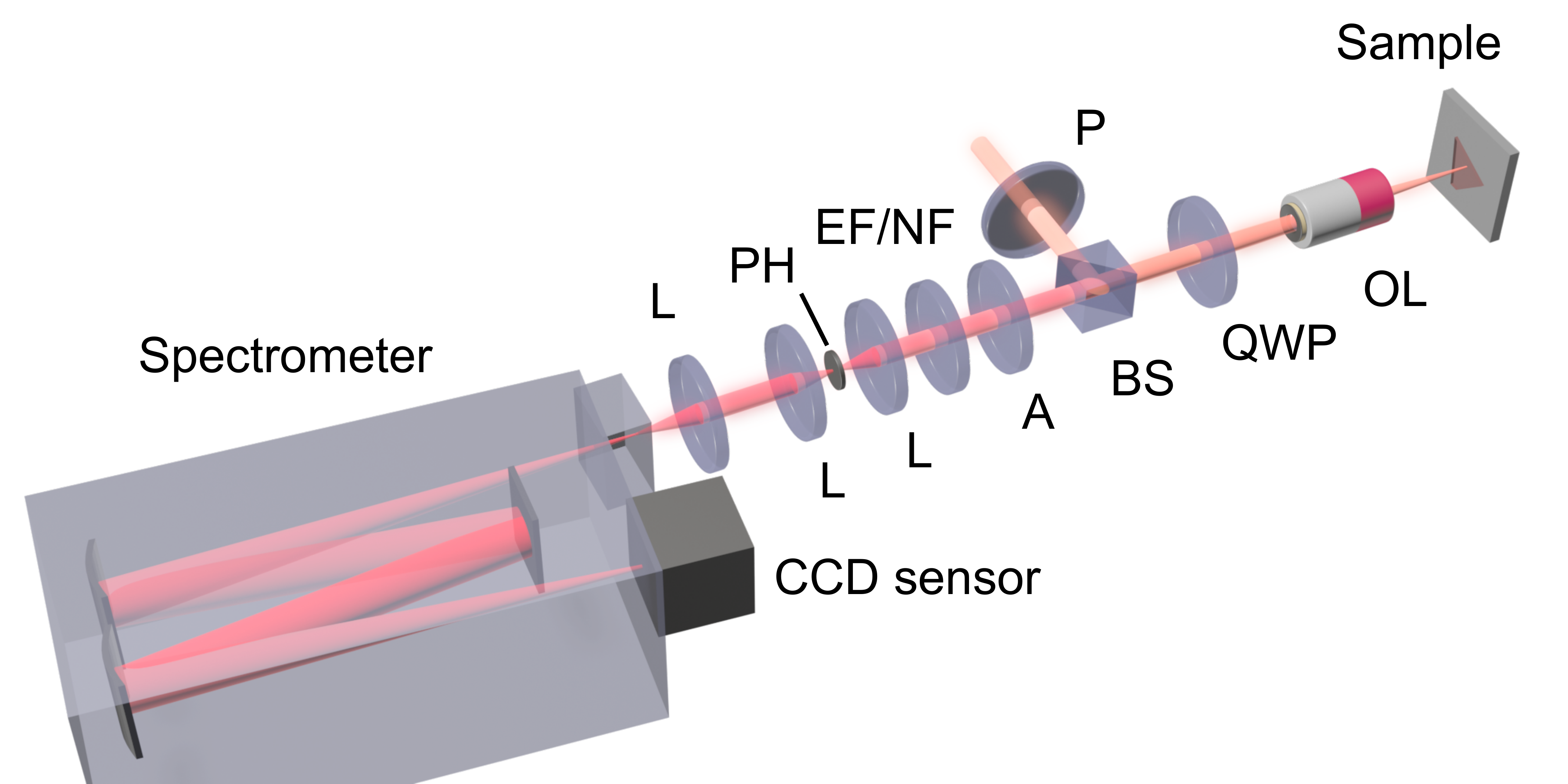}
  \caption{\label{fig:setup}
  \addr{Optical setup of circularly polarized Raman spectroscopy. P: polarizer, A: analyzer, BS: beam splitter, OL: objective lens, QWP: quarter wave-plate, EF/NF: edge/notch filter, PH: pinhole, L: lens.}
  }
\end{figure}

\clearpage
\section{\label{app:tens}\NoCaseChange{\addb{Raman tensors for circular polarization configurations}}}
The Raman intensity in the electric dipole approximation for given incident and scattered polarization vectors, $\vbe{i}$ and $\vbe{s}$ respectively, is expressed as
$I\propto \qty|\vbe{s}^\dagger R\vbe{i}|^2$.
The representation $\Gamma$ of the phonon mode excited in the Stokes process is determined as $\Gamma_{\PH}=\Gamma_{\vbe{s}^*}\otimes\Gamma_{\vbe{i}}$. 
In the point group $\bar{3}$ of \ce{NiTiO3}, the polarization vectors $\vbe{R}$ and $\vbe{L}$ of circulary polarized light propagating along the $z$-axis correspond to the irreducible representations $\nEu{1}$ and $\nEu{2}$, respectively.
Consequently, the phonon modes excited in the cross-circular polarization configurations are
\begin{align}
  \GMRL_{\PH}&=\nEu{1}\otimes\nEu{1}=\nEg{2}, \\
  \GMLR_{\PH}&=\nEu{2}\otimes\nEu{2}=\nEg{1}.
\end{align}
These selection rules are consistent with those obtained from the \eraseb{\addb{PAM}}\addb{crystal (pseudo)-angular momentum (CAM)} conservation law~\cite{tatsumi2018,ishito2023a}.

The Raman tensor for the $\nEg{i}~(i=1,\pad 2)$ representation is derived by using the projection operator method.
The projection operator is given by~\cite{dresselhaus2007}
\begin{align}
  P^{\Gamma_{n}}_{\mu\nu}=\frac{d_{n}}{h}\sum_{R}D^{\Gamma_{n}}_{\mu\nu}(R)^*R,
  \label{eq:proj}
\end{align}
where $d_{n}$ denotes the dimensionality of the irreducible representation $\Gamma_{n}$,
$h$ denotes the number of symmetry operators in the group, 
and $D^{\Gamma_{n}}_{\mu\nu}(R)$ is the matrix representation of the symmetry operator $R$.
By constructing the projection operator for the $\nEg{1}$-irreducible representation using the character table (Table~\ref{tbl:character_3i}) and applying it to the components of the rank-2 polar tensor, we obtain
\begin{align}
  &\quad P_{11}^{\nEg{1}}xx \notag \\
  &=\frac{1}{6}
  \begin{aligned}[t]
      &\left[1\cdot xx+w^{2*}\cdot\qty(-\frac12x+\frac{\sqrt3}{2}y)\qty(-\frac12x+\frac{\sqrt3}{2}y)
      +w^{*}\cdot\qty(-\frac12x-\frac{\sqrt3}{2}y)\qty(-\frac12x-\frac{\sqrt3}{2}y)\right. \notag \\
      &+\left.1\cdot\qty(-x)\qty(-x)+w^{2*}\cdot\qty(\frac12x-\frac{\sqrt3}{2}y)\qty(\frac12x-\frac{\sqrt3}{2}y)
      +w^{*}\cdot\qty(\frac12x+\frac{\sqrt3}{2}y)\qty(\frac12x+\frac{\sqrt3}{2}y)\right]
  \end{aligned} \notag \\
  &=\frac14\qty(xx-yy)-i\frac14\qty(xy+yx), 
\end{align}
\begin{align}
  &\quad P_{11}^{\nEg{1}}yy \notag \\
  &=\frac{1}{6}
  \begin{aligned}[t]
      &\left[1\cdot yy+w^{2*}\cdot\qty(-\frac{\sqrt3}{2}x-\frac12y)\qty(-\frac{\sqrt3}{2}x-\frac12y)
      +w^{*}\cdot\qty(\frac{\sqrt3}{2}x-\frac12y)\qty(\frac{\sqrt3}{2}x-\frac12y)\right. \notag \\
      &\left.+1\cdot\qty(-y)\qty(-y)+w^{2*}\cdot\qty(\frac{\sqrt3}{2}x+\frac12y)\qty(\frac{\sqrt3}{2}x+\frac12y)
      +w^{*}\cdot\qty(-\frac{\sqrt3}{2}x+\frac12y)\qty(-\frac{\sqrt3}{2}x+\frac12y)\right]
  \end{aligned} \notag \\
  &=-\frac14\qty(xx-yy)+i\frac14\qty(xy+yx), 
\end{align}
\begin{align}
  &\quad P_{11}^{\nEg{1}}xy \notag \\
  &=\frac{1}{6}
  \begin{aligned}[t]
      &\left[1\cdot xy+w^{2*}\cdot\qty(-\frac12x+\frac{\sqrt3}{2}y)\qty(-\frac{\sqrt3}{2}x-\frac12y)
      +w^{*}\cdot\qty(-\frac12x-\frac{\sqrt3}{2}y)\qty(\frac{\sqrt3}{2}x-\frac12y)\right. \notag \\
      &\left.+1\cdot\qty(-x)\qty(-y)+w^{2*}\cdot\qty(\frac12x-\frac{\sqrt3}{2}y)\qty(\frac{\sqrt3}{2}x+\frac12y)
      +w^{*}\cdot\qty(\frac12x+\frac{\sqrt3}{2}y)\qty(-\frac{\sqrt3}{2}x+\frac12y)\right]
  \end{aligned} \notag \\
  &=-i\frac14\qty(xx-yy)-\frac14\qty(xy+yx), 
\end{align}
and similarly $P_{11}^{\nEg{1}}yx=-i\frac14\qty(xx-yy)-\frac14\qty(xy+yx)$.
Thus, the rank-2 polar tensor for the $\nEg{1}$ representation becomes
\begin{align}
    P_{11}^{\nEg{1}}R_{ij}&=\mqty(\frac14\qty(xx-yy)-i\frac14\qty(xy+yx) & -i\frac14\qty(xx-yy)-\frac14\qty(xy+yx) \\
    -i\frac14\qty(xx-yy)-\frac14\qty(xy+yx) & -\frac14\qty(xx-yy)+i\frac14\qty(xy+yx)) \notag \\
    &=\frac14\qty[\qty(xx-yy)-i\qty(xy+yx)]
    \mqty(1 & -i \\
    -i & 1).
\end{align}
\addb{Using an analogous procedure, 
the rank-2 polar tensor for the $\nEg{2}$ representation becomes
\begin{align}
    P_{11}^{\nEg{2}}R_{ij}&=\mqty(\frac14\qty(xx-yy)+i\frac14\qty(xy+yx) & i\frac14\qty(xx-yy)-\frac14\qty(xy+yx) \\
    i\frac14\qty(xx-yy)-\frac14\qty(xy+yx) & -\frac14\qty(xx-yy)-i\frac14\qty(xy+yx)) \notag \\
    &=\frac14\qty[\qty(xx-yy)+i\qty(xy+yx)]
    \mqty(1 & i \\
    i & 1).
\end{align}
}

Therefore, 
\addb{the Raman tensor for the LR configuration (projecting to the $\nEg{2}$ representation) with an excitation of the $\nEg{1}$ phonon is given by
\begin{align}
  R^{\mathrm{LR}}=
  R_{1}\mqty(1 & i \\
  i & -1),
    \label{LR_raman_tensor}
\end{align}
where $R_{1}$ is a proportional coefficient associated with the $\nEg{1}$ phonons.
}

\eraseb{\addb{the Raman tensor for the RL configuration (projecting to the $\nEg{1}$ representation) with an excitation of the $\nEg{2}$ phonon can be expressed as}}
\eraseb{\addb{where $R_{2}$ is a proportional coefficient associated with the $\nEg{2}$ phonons. }}

\eraseb{\addb{Using an analogous procedure, }}
\addb{The Raman tensor for the RL configuration (projecting to the $\nEg{1}$ representation) with an excitation of the $\nEg{2}$ phonon can be expressed as
\begin{align}
  R^{\mathrm{RL}}=
    R_{2}\mqty(1 & -i \\
    -i & -1),
    \label{RL_raman_tensor}
\end{align}
where $R_{2}$ is a proportional coefficient associated with the $\nEg{2}$ phonons. 
}

\eraseb{\addb{the Raman tensor for the LR configuration (projecting to the $\nEg{2}$ representation) with an excitation of the $\nEg{1}$ phonon is given by}}
\eraseb{\addb{where $R_{1}$ is a proportional coefficient associated with the $\nEg{1}$ phonons.}}

\addr{We can check the selection rules for circular polarization configurations
using the circular polarization basis $\vbe{R}=\transp{(1/\sqrt{2},\pad i/\sqrt{2})}$, $\vbe{L}=\transp{(1/\sqrt{2},\pad -i/\sqrt{2})}$.
The intensities derived from the Raman tensor in Eqs.~\eqref{LR_raman_tensor} and \eqref{RL_raman_tensor} yield:
\begin{align}
  \ILR&\propto\qty|\frac{R_{1}}{2}\qty(1,\pad -i)\mqty(1 & i \\ i & -1)\mqty(1 \\ -i)|^2
  +\qty|\frac{R_{2}}{2}\qty(1,\pad -i)\mqty(1 & -i \\ -i & -1)\mqty(1 \\ -i)|^2
  =2\qty|R_{1}|^2, \\
  \IRL&\propto\qty|\frac{R_{1}}{2}\qty(1,\pad i)\mqty(1 & i \\ i & -1)\mqty(1 \\ i)|^2
  +\qty|\frac{R_{2}}{2}\qty(1,\pad i)\mqty(1 & -i \\ -i & -1)\mqty(1 \\ i)|^2
  =2\qty|R_{2}|^2, \\
  \ILL&\propto\qty|\frac{R_{1}}{2}\qty(1,\pad i)\mqty(1 & i \\ i & -1)\mqty(1 \\ -i)|^2
  +\qty|\frac{R_{2}}{2}\qty(1,\pad i)\mqty(1 & -i \\ -i & -1)\mqty(1 \\ -i)|^2
  =0, \\
  \IRR&\propto\qty|\frac{R_{1}}{2}\qty(1,\pad -i)\mqty(1 & i \\ i & -1)\mqty(1 \\ i)|^2
  +\qty|\frac{R_{2}}{2}\qty(1,\pad -i)\mqty(1 & -i \\ -i & -1)\mqty(1 \\ i)|^2
  =0.
\end{align}
Here, we assume that the intensities associated with $R_{1}$ and $R_{2}$ are independent.
Therefore, only cross-circular polarzation (LR and RL) configurations give finite intensities 
and ROA appears when $\qty|R_{1}|\neq\qty|R_{2}|$, consistent with our experimental observations.
}

\addr{
In addition, the phonon modes excited in the parallel-circular polarization configurations are
\begin{align}
  \GMLL_{\PH}&=\nEu{1}\otimes\nEu{2}=\Ag, \\
  \GMRR_{\PH}&=\nEu{2}\otimes\nEu{1}=\Ag.
\end{align}
This selection rules are confirmed by using the Raman tensor for $\Ag$ modes,
\begin{align}
  R\propto\mqty(1 & a \\ -a & 1),
\end{align}
as follows:
\begin{align}
  \ILR&\propto\qty|\frac{1}{2}\qty(1,\pad -i)\mqty(1 & a \\ -a & 1)\mqty(1 \\ -i)|^2
  =0, \\
  \IRL&\propto\qty|\frac{1}{2}\qty(1,\pad i)\mqty(1 & a \\ -a & 1)\mqty(1 \\ i)|^2
  =0, \\
  \ILL&\propto\qty|\frac{1}{2}\qty(1,\pad i)\mqty(1 & a \\ -a & 1)\mqty(1 \\ -i)|^2
  =\qty|1-ai|^2, \\
  \IRR&\propto\qty|\frac{1}{2}\qty(1,\pad -i)\mqty(1 & a \\ -a & 1)\mqty(1 \\ i)|^2
  =\qty|1+ai|^2.
\end{align}
Therefore, only parallel-circular polarization (LL and RR) configurations \eraseb{\addb{(LL and RR) }}give finite intensities for the $\Ag$ modes.
Moreover, we can assume that the antisymmetric component $a$, which is the source of ROA in parallel-circular polarization configurations, is negligibly small ($|a|\approx0$), because no ROA was observed for the $\Ag$ modes (Sec.~\ref{app:ag}).
}

\addr{
In general, Raman intensity is given by \cite{jorio2011}
\begin{align}
  I\propto\qty|\sum_{m,m^\prime}\frac{\bra{i}\Hopt\ket{m^\prime}\bra{m^\prime}\Hep\ket{m}\bra{m}\Hopt\ket{i}}{(\hbar\omega_{m^\prime i}+\hbar\omega-\hbar\omega_{\mathrm{i}}-i\gamma)(\hbar\omega_{m i}-\hbar\omega_{\mathrm{i}}-i\gamma)}|^2,
\end{align}
where $\Hopt$ is the electron-photon interaction Hamiltonian, $\Hep$ is the electron-phonon interaction Hamiltonian. 
$\hbar\omega$, $\hbar\omega_{\mathrm{i}}$, and $\hbar\omega_{\mathrm{s}}$ are the energy of phonon, incident photon, and scattered photon, respectively. 
$\hbar\omega_{m^{(\prime)}i}$ is the energy difference between electron's initial state $\ket{i}$ and intermediate state $\ket{m^{(\prime)}}$. $\gamma$ is the damping coefficient. 
}

\addr{
Because the actual site symmetry of \ce{Ni^{2+}} ions is point group $3$, the Raman intensity is strongly enhanced at resonance via electric-dipole-allowed transitions described by $\Hopt=-\vb{p}\vdot\vb{E}$, 
particularly near the \ce{Ni^{2+}} ${}^3A_{2\mathrm{g}}\to {}^3T_{1\mathrm{g}}$ transition at \Sqty{785}{nm}, where $\vb{p}$ and $\vb{E}$ are the electric dipole moment and the electric field of light, respectively. 
This resonance effect significantly increases the Raman susceptibility and thus the observed ROA intensity.
}


\addr{
Importantly, while the resonance strongly modifies the magnitude and phase of the Raman tensor elements, it does not change the symmetry classification of the Raman-active phonons or the polarization selection rules that determine which irreducible representations are allowed.
The selection rules themselves are determined by the electron-phonon matrix element $\bra{m^\prime}\Hep\ket{m}$, which is constrained solely by crystal symmetry, in the same manner as in the off-resonant case. 
We also confirmed that our mode assignment at \Sqty{785}{nm} is consistent with the ordinary selection rules [Fig.~2(c)]. 
Under resonant conditions, the Raman intensity of $\Eg$ modes may generally contain an interference term between the two degenerate components.
However, such interference cannot give rise to ROA because $\nEg{1}$ and $\nEg{2}$ modes are eigenmodes of the cross-circular polarization configurations.}

\clearpage
\section{\label{app:fpc}\NoCaseChange{Properties of phonons in \ce{NiTiO3}}}
To investigate the phonon modes excited by the circularly polarized Raman scattering, 
\addb{we performed first-priciples calculations.}
\eraseb{\addb{d}}\addb{D}ensity functional theory (DFT) calculations were \eraseb{\addb{performed}}\addb{carried out} 
using the projector augmented-wave (PAW) method~\cite{blochl1994}, 
as implemented in the VASP code~\cite{kresse1996,kresse1999}. 
The radial cutoffs in the PAW data-sets are 1.29, 1.32, and \Sqty{0.82}{\angstrom} for \ce{Ni}, \ce{Ti}, and \ce{O}, respectively. 
A cutoff energy of \Sqty{550}{eV} was used for the plane waves. 
All the calculations were performed using the Perdew--Burke--Ernzerhof form of the generalized gradient approximation 
optimized for solids (GGA-PBEsol) to treat exchange--correlation interactions~\cite{perdew1996}. 
The following states were regarded as the valence states: $3d$ and $4s$ for \ce{Ni}, $3d$ and $4s$ for \ce{Ti}, and $2s$ and $2p$ for \ce{O}. 
We employed a $7\times7\times7$ $\Gamma$-centered $k$-point mesh. 
To account for strongly correlated $d$ electrons, 
the Hubbard $U$ term~\cite{dudarev1998} for $U_{\mathrm{eff}} = \Sqty{6}{eV}$ for \ce{Ni} was included~\cite{zhou2004a}. 
The lattice constants and internal coordinates 
were optimized until the residual stress and forces converged down to \Sqty{0.01}{GPa} and \Sqty{10}{meV/\angstrom}, respectively. 
The phonon band structures were derived from the calculated force constants using the Phonopy code~\cite{togo2015}. 
A $2\times2\times2$ supercell was utilized to calculate the the phonon frequencies.

The resulting phonon frequencies and corresponding modes are listed in Table~\ref{tbl:phonon_freq}.
\addr{Here, the $\Ag^{(5)\prime}$ mode, which can be observed only in experiments, is considered to come from the amorphous phase of \ce{NiTiO3}~\cite{ruizpreciado2015,tursun2019,qi2022}.
}
\eraseb{\addb{, and t}}\addb{T}he phonon band structure is shown in Fig.~\subref{fig:phononband}{a}.
\addr{Each displacement vector of atoms in the unit cell for all $\Eg$ modes are depicted in Figs.~\subref{fig:phono_eigendips}{a}--\subref{fig:phono_eigendips}{e}.
}

Figures~\subref{fig:phononband}{b} and \subref{fig:phononband}{c} show the calculated phonon angular momentum~\cite{zhang2014}, categorized by the irreducible representations of point group $3$ along the $\Gamma$--$Z$ path.
The character table for point group $3$ is shown in Table~\ref{tbl:character_3}. 
At the $\Gamma$ point, the point group is $\bar{3}$, and the character table is shown in Table~\ref{tbl:character_3i}.
Owing to the presence of spatial inversion and time-reversal symmetry in \ce{NiTiO3}, the doubly degenerate phonon modes at the $\Gamma$ point remain degenerate along the $\Gamma$--$Z$ path.
Although the individual ${}^1E$ and ${}^2E$ phonons may possess positive or negative angular momentum, their degeneracy results in a net angular momentum of zero at each point along $\Gamma$--$Z$.
However, based on the \eraseb{\addb{pseudoangular momentum (PAM)}}\addb{CAM} conservation law, the Stokes processes in \eraseb{\addb{RL}}\addb{LR} and \eraseb{\addb{LR}}\addb{RL} configurations can selectively excite phonons corresponding to the irreducible representations \eraseb{\addb{${}^2E$}}\addb{${}^1E$} and \eraseb{\addb{${}^1E$}}\addb{${}^2E$}, respectively.
Thus, the \eraseb{\addb{RL}}\addb{LR} and \eraseb{\addb{LR}}\addb{RL} configurations excite phonons with opposite signs of \eraseb{\addb{PAM}}\addb{CAM}.
These phonons along the $\Gamma$--$Z$ path possess nonzero wavenumbers and angular momentum, qualifying them as truly chiral phonons~\cite{ishito2023a}.
It is important to note that the chiral phonons discussed herein are distinct from those in chiral crystals such as $\alpha$-\ce{HgS}~\cite{ishito2023a}, \ce{Te}~\cite{ishito2023c}, and $\alpha$-quartz~\cite{oishi2024}.
In our case, there is no linear band splitting at finite wavenumbers, no band swapping at the Brillouin zone boundary, and no phonon group velocity at the $\Gamma$ point.

\begin{table}[H]
  \centering
  \caption{\label{tbl:phonon_freq}
  Frequencies and symmetries of phonon modes at the $\Gamma$ point in \ce{NiTiO3}, calculated using VASP and Phonopy. 
  Experimental values from this work and Ref.~\cite{qi2022} are also provided for comparison.}
  \begin{ruledtabular}
    \begin{tabular}{cccccccccccc}
      & {$\Ag^{(1)}$} & {$\Eg^{(1)}$} & {$\Ag^{(2)}$} & {$\Eg^{(2)}$} & {$\Eg^{(3)}$} 
      & {$\Ag^{(3)}$} & {$\Eg^{(4)}$} & {$\Ag^{(4)}$} & {$\Eg^{(5)}$} & {$\Ag^{(5)}$} & {\addr{$\Ag^{(5)\prime}$}} \\
      \colrule
      Cal. (\unit{cm^{-1}}) & 195 & 224 & 242 & 288 
      & 332 & 387 & 453 & 473 & 626 & 701 & \addr{-} \\
      Exp. (\unit{cm^{-1}}) & 196 & 232 & 250 & 294 
      & 348 & 398 & 468 & - & 615 & 709 & \addr{738} \\
      Ref.~\cite{qi2022} (\unit{cm^{-1}}) & 195 & 232 & 250 & 295 
        & 350 & 398 & 469 & 488 & 620 & 716 & \addr{773} \\
    \end{tabular}
  \end{ruledtabular}
\end{table}
\begin{figure}[H]
  \centering
  \includegraphics[width=0.9\hsize]{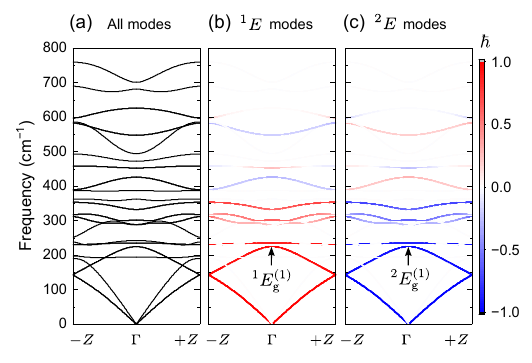}
  \caption{\label{fig:phononband}
  Phonon dispersion and angular momentum of \ce{NiTiO3} calculated via first-principles methods.
  (a) Phonon dispersion along the $-Z~(-0.5,-0.5,-0.5)$--$\Gamma~(0.0,0.0,0.0)$--$+Z~(0.5,0.5,0.5)$ path.
  (b), (c) $z$-components of the phonon angular momentum for the \eraseb{\addb{${}^2E$}}\addb{${}^1E$} and \eraseb{\addb{${}^1E$}}\addb{${}^2E$} irreducible representations, respectively.
  \eraseb{\addb{(d) Atomic displacement pattern of the $\Eg^{(1)}$ mode at the $\Gamma$ point.}}
  }
\end{figure}
\begin{figure}[H]
  \centering
  \includegraphics[width=0.9\hsize]{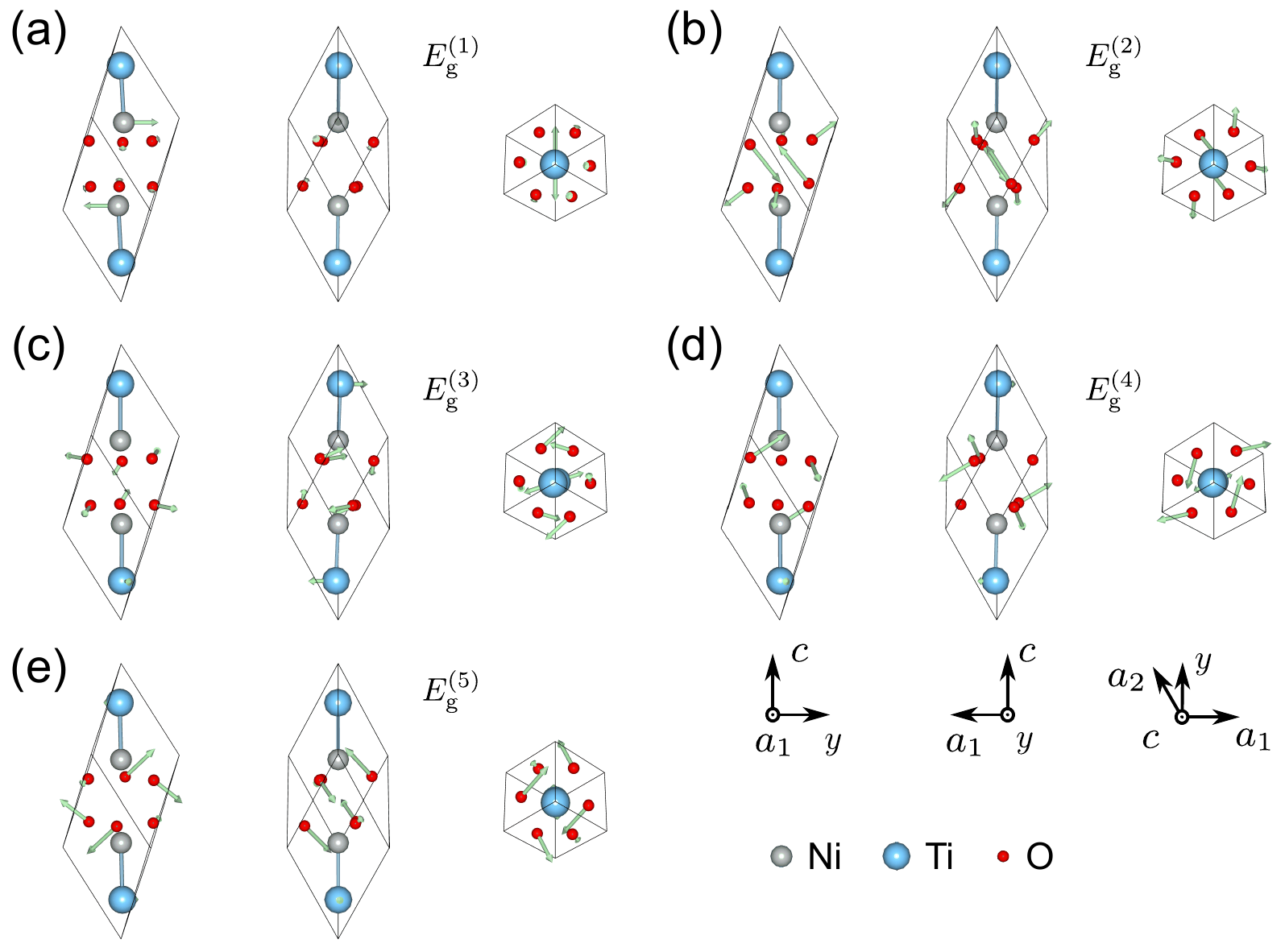}
  \caption{\label{fig:phono_eigendips}
  \addr{Eigendisplacements of the five $\Eg$ phonon modes calculated by the first-principles calculation. (a) $\Eg^{(1)}$, (b) $\Eg^{(2)}$, (c) $\Eg^{(3)}$, (d) $\Eg^{(4)}$, and (e) $\Eg^{(5)}$ modes, respectively.
  Only one component of the doubly degenerate $\Eg$ modes is visualized.}
  }
\end{figure}
\begin{table}[H]
  \caption{\label{tbl:character_3}Character table of point group $3$. $w=\exp(2\pi i/3)$.}
  \begin{ruledtabular}
    \begin{tabular}{crcccc}
      & & $E$ & $C_3^+$ & $C_3^-$ & \\
      \colrule
      & $A$ & 1 & 1 & 1 & \\
      & ${}^1E$ & 1 & $w^2$ & $w$ & \\
      & ${}^2E$ & 1 & $w$ & $w^2$ & \\
    \end{tabular}
  \end{ruledtabular}
\end{table}
\begin{table}[H]
  \caption{\label{tbl:character_3i}Character table of point group $\bar{3}$. $w=\exp(2\pi i/3)$.}
  \begin{ruledtabular}
    \begin{tabular}{crccccccc}
      & & \multicolumn{1}{c}{$E$} & \multicolumn{1}{c}{$C_3^+$} & \multicolumn{1}{c}{$C_3^-$}
      & \multicolumn{1}{c}{$i$} & \multicolumn{1}{c}{$C_{3i}^+$} & \multicolumn{1}{c}{$C_{3i}^-$} & \\
      \colrule
      & $\Ag$ & \num{1} & \num{1} & \num{1} & \num{1} & \num{1} & \num{1} & \\
      & $\nEg{1}$ & \num{1} & \multicolumn{1}{c}{$w^2$} & \multicolumn{1}{c}{$w$} 
      & \num{1} & \multicolumn{1}{c}{$w^2$} & \multicolumn{1}{c}{$w$} & \\
      & $\nEg{2}$ & \num{1} & \multicolumn{1}{c}{$w$} & \multicolumn{1}{c}{$w^2$} 
      & \num{1} & \multicolumn{1}{c}{$w$} & \multicolumn{1}{c}{$w^2$} & \\
      & $\Au$ & \num{1} & \num{1} & \num{1} & \num{-1} & \num{-1} & \num{-1} & \\
      & $\nEu{1}$ & \num{1} & \multicolumn{1}{c}{$w^2$} & \multicolumn{1}{c}{$w$} 
      & \num{-1} & \multicolumn{1}{c}{$-w^2$} & \multicolumn{1}{c}{$-w$} & \\
      & $\nEu{2}$ & \num{1} & \multicolumn{1}{c}{$w$} & \multicolumn{1}{c}{$w^2$} 
      & \num{-1} & \multicolumn{1}{c}{$-w$} & \multicolumn{1}{c}{$-w^2$} & \\
  \end{tabular}
  \end{ruledtabular}
\end{table}

\clearpage
\section{\label{app:wav}\NoCaseChange{Excitation wavelength dependence for cross-circular ROA}}
Raman spectra in cross-circular polarization configurations obtained using an excitation wavelength of \Sqty{532}{nm} \addr{and \Sqty{633}{nm}} are shown in Figs.~\ref{fig:532_spectrum} \addr{and \ref{fig:633_spectrum}, respectively}.
As with the \Sqty{785}{nm} excitation, five peaks corresponding to $\Eg$ phonon modes were observed.
\eraseb{\addb{However, the circular intensity difference $\groa$ for each peak was found to be nearly zero.}}
The intensity differences in the Raman spectra in the cross-circular \addb{polarization} configurations defined as {$\groa=2(\ILR-\IRL)/(\ILR+\IRL)$} are listed in Table~\ref{tbl:rho}.
\addr{Although much weaker, we indeed observe discernible ROA signals at \Sqty{532}{nm}, while ROA was negligibly small at \Sqty{633}{nm}.}
This result indicates that the ROA of \ce{NiTiO3} is strongly enhanced by the resonance of the excitation wavelength.



\begin{figure}[H]
    \centering
    \includegraphics[width=0.7\hsize]{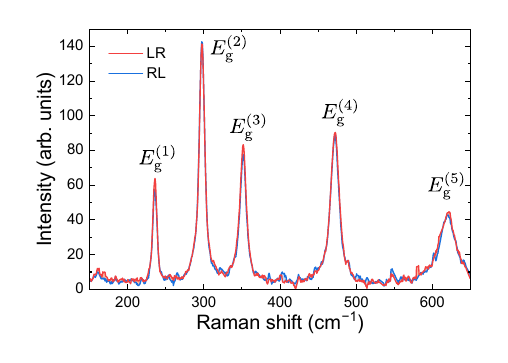}
    \caption{\label{fig:532_spectrum}
    Raman spectra of the front surface of a single-domain \ce{NiTiO3} crystal measured in the cross-circular polarization configurations at \Sqty{295}{K}, using an excitation wavelength of \Sqty{532}{nm}.}
\end{figure}
\begin{figure}[H]
    \centering
    \includegraphics[width=0.7\hsize]{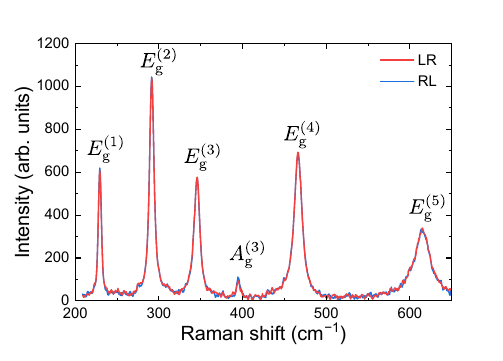}
    \caption{\label{fig:633_spectrum}
    \addr{Raman spectra of the front surface of a single-domain \ce{NiTiO3} crystal measured in the cross-circular polarization configurations at \Sqty{295}{K}, using an excitation wavelength of \Sqty{633}{nm}.}}
\end{figure}

\begin{table}[H]
  \centering
  \caption{The values of $\groa$ for five $\Eg$ peaks excited with wavelengths of \Sqty{785}{nm}\addr{,}\eraser{\addr{and}} \Sqty{532}{nm}\addr{, and \Sqty{633}{nm}} for the single-domain crystal. 
  The error is \num{+-0.02} for each peak. \label{tbl:rho}}
  \begin{ruledtabular}
    \begin{tabular}{cSSSSS}
      & {$\Eg^{(1)}$} & {$\Eg^{(2)}$} & {$\Eg^{(3)}$} & {$\Eg^{(4)}$} & {$\Eg^{(5)}$} \\
      \colrule
      $\groa$ (\Sqty{785}{\nm}, front) & 0.96 & -0.16 & 0.21 & -0.01 & 0.47 \\
      $\groa$ (\Sqty{785}{\nm}, back) & -1.04 & 0.11 & -0.27 & -0.05 & -0.47 \\
      \colrule
      $\groa$ (\Sqty{532}{\nm}, front) & 0.10 & 0.01 & 0.10 & 0.05 & 0.09 \\
      \colrule
      \addr{$\groa$ (\Sqty{633}{\nm}, front)} & \addr{0.00} & \addr{0.00} & \addr{0.02} & \addr{-0.01} & \addr{0.02}
  \end{tabular}
  \end{ruledtabular}
\end{table}

\clearpage
\section{\NoCaseChange{\addr{Differential spectrum}}}
\addr{Differential spectra corresponding to Fig.~\eraseb{\addb{2}}\addb{3} of the main text is shown in Figs.~\subref{fig:diff_785}{a} and \subref{fig:diff_785}{b}.
Clear intensity differences are observed for $\Eg$ modes, revealing a sign reversal between the front and back surfaces of the crystal.}

\begin{figure}[H]
    \centering
    \includegraphics[width=0.98\hsize]{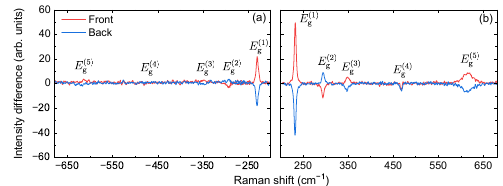}
    \caption{\label{fig:diff_785}
    \addr{Differential Raman spectra of single-crystal \ce{NiTiO3} measured from the front and back surfaces of the crystal with excitation wavelength of \Sqty{785}{nm}.
    (a) anti-Stokes and (b) Stokes spectra, respectively.}}
\end{figure}

\clearpage
\section{\label{app:ag}\NoCaseChange{\addb{Raman measurements in parallel-circular polarization configurations}}}
Raman spectra measured in parallel-circular polarization \addb{(LL and RR)} configurations \eraseb{\addb{(LL and RR) }}are shown in Fig.~\ref{fig:PCROA}.
For the four observed $A_g$ modes, no intensity difference was found between the two configurations.

\begin{figure}[H]
  \centering
  \includegraphics[width=0.7\hsize]{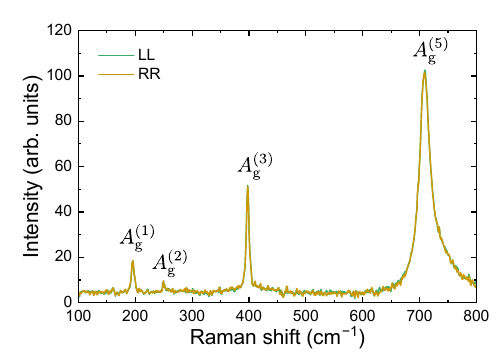}
  \caption{\label{fig:PCROA}
  Raman spectra of the single-domain \ce{NiTiO3} crystal obtained in parallel-circular polarization configurations at \Sqty{295}{K}, using an excitation wavelength of \Sqty{785}{nm}.
  }
\end{figure}

\clearpage
\section{\label{app:mdl}\NoCaseChange{Details of model calculation}}
In the electric dipole approximation, Raman scattering can be expressed as
\begin{align}
  P_{i}(\omega+\delta\omega)
  =\chi_{ij}^{\PH}(\omega,\delta\omega)\PH(\delta\omega)E_{j}(\omega),
\end{align}
where $P_i(\omega+\delta\omega)$ is the induced polarization, $E_j(\omega)$ is the electric field,
$\PH(\delta\omega)$ denotes the normal coordinate of phonons with frequency $\delta\omega$, and $\chi_{ij}^{\PH}(\omega,\delta\omega)$ is the second-order nonlinear susceptibility tensor.
A negative (positive) sign of $\delta \omega$ indicates the Stokes (anti-Stokes) process.
The Raman tensor, i.e., the electric susceptibility modified by a given phonon excitation $\PH$, is obtained as
\begin{equation}
   R_{ij} (\omega, \delta\omega)
  =\chi_{ij}^{\PH}(\omega,\delta\omega)\PH(\delta\omega).
\end{equation}

$\PH_{1}(\delta\omega)$ and $\PH_{2}(\delta\omega)$ represent the phonon excitations belonging to the irreducible representations $\nEg{1}$ and $\nEg{2}$ of the point group $\bar{3}$, respectively.
Owing to the product rule of the irreducible representations, the nonlinear susceptibilities $\chi_{ij}^{\PH_1}$ and $\chi_{ij}^{\PH_2}$ for the phonon excitations $\PH_1$ and $\PH_2$ belong to the $\nEg{2}$ and $\nEg{1}$ representations, respectively.
The symmetry-adapted forms can be obtained as
\begin{align}
    \hat{\chi}^{\PH_1} 
    = \chi_1 \begin{pmatrix}
        1 & i & 0 \\
        i & -1& 0\\
        0&0&0
    \end{pmatrix}, ~
    \hat{\chi}^{\PH_2} 
    = \chi_2 \begin{pmatrix}
        1 & -i &0\\
        -i & -1 &0\\
        0&0&0
    \end{pmatrix}.
\end{align}
Note that the obtained bilinear forms of nonlinear susceptibility and phonon intensity are related to Eqs.~\eqref{LR_raman_tensor} and \eqref{RL_raman_tensor} as 
\begin{equation}
  R_{1} = \chi_1 \PH_1, ~ R_{2} = \chi_2 \PH_2,
\end{equation}
yielding cross-circular Raman scattering intensities proportional to
\begin{align}
    \qty|R_{1}|^2 = |\chi_1|^2 |\PH_1|^2,
\end{align}
for the LR configuration, and 
\begin{align}
    \qty|R_{2}|^2 = |\chi_2|^2 |\PH_2|^2,
\end{align}
for the RL configuration.
As a result, the cross-circular ROA, evaluated by $\qty|R_{1}|^2-\qty|R_{2}|^2 = |\chi_1|^2 |\PH_1|^2 - |\chi_2|^2 |\PH_2|^2$, originates from the imbalance between $|\chi_1|$ and $|\chi_2|$, as well as between $|\PH_1|$ and $|\PH_2|$.

Given that the phonon excitations are real, the phonon modes of interest, $\PH_i (\delta \omega)$, satisfy the relation
\begin{equation}
    \PH_1 (\delta \omega) = \PH_2 (-\delta \omega).
\end{equation}
Furthermore, time-reversal symmetry ensures the relation
\begin{equation}
    \PH_1 (\delta \omega) = \PH_1 (-\delta \omega),
\end{equation}
which yields the relationship $\PH_1 (\delta \omega) = \PH_2 (\delta \omega)$.
As a result, the cross-circular ROA of the ferroaxial system is determined by the difference in the nonlinear susceptibilities.
Using the normalized values
\begin{align}
    \overline{I}_\text{LR} \equiv \frac{|\chi_1 \PH_1|^2 }{|\chi_1 \PH_1|^2 + |\chi_2 \PH_2|^2}
        & = \frac{|\chi_1 |^2 }{|\chi_1 |^2 + |\chi_2 |^2},
        \label{normalized_LR_intensity}\\
    \overline{I}_\text{RL} \equiv \frac{|\chi_2 \PH_2|^2 }{|\chi_1 \PH_1|^2 + |\chi_2 \PH_2|^2}
        & = \frac{|\chi_2 |^2 }{|\chi_1 |^2 + |\chi_2 |^2},\label{normalized_RL_intensity}
\end{align}
we evaluate the cross-circular ROA as follows:
\begin{equation}
    \frac{I_\text{LR} - I_\text{RL}}{I_\text{LR} + I_\text{RL}} \propto  \frac{|\chi_1|^2 - |\chi_2 |^2}{|\chi_1 |^2 + |\chi_2 |^2}.
    \label{cross_circular_ROA}
\end{equation}

We introduced the model Hamiltonian to calculate the cross-circular ROA of ferroaxial materials.
The tight-binding Hamiltonian defined in the triangular lattice is
\begin{equation}
    H = \sum_\bk \bc_\bk^\dagger H_\bk \bc_\bk,
\end{equation}
where the creation operators in the vector $\bc_\bk^\dagger$ are parameterized by the crystal momentum $\bk$, sublattice degrees of freedom ($\alpha,\beta$), and $p$-orbital degrees of freedom ($p_x,p_y,p_z$) as
\begin{equation}
    \bc_\bk^\dagger = (c_{\bk \alpha x}^\dagger,c_{\bk \alpha y}^\dagger,c_{\bk \alpha  z}^\dagger, c_{\bk \beta x}^\dagger,c_{\bk \beta y}^\dagger,c_{\bk \beta  z}^\dagger).
\end{equation}
The site symmetry of each sublattice is described by the chiral and polar point group $3$.
The matrix elements of $H_\bk$ are expressed in the symmetry-adapted form of the ferroaxial point group $\bar{3}$.
When the sublattice space is expressed by Pauli matrices $\tau_a$, the Hamiltonian is
                \begin{equation}
                H_\bk = H_\bk^1 + H_\bk^2,
                \end{equation}
where each term is obtained by a multipole-based symmetry analysis~\cite{kusunose2023} as 
                \begin{align}
                H_\bk^1 &= 
                t_1 \tau_0 \otimes 
                        \begin{pmatrix}
                                 1& 0&0 \\
                                 0& 1&0 \\
                                 0&0 &-2 
                        \end{pmatrix}\notag \\
                &+ t_2 \left(\cos{q_z}\tau_x  - \sin{q_z}\tau_y \right)\otimes \bm{1}_3  \notag \\
                &+ t_3 \left(\cos{q_x} + \cos{q_y} + \cos{(q_x+q_y)}\right) \bm{1}_6 \notag \\
                &+ \frac{\sqrt{6}}{12}t_4 \tau_0 \otimes 
                \begin{pmatrix}
                        0&0 &h_1 (\bq) \\
                        0&0 &h_2 (\bq) \\
                         h_1 (\bq)&h_2 (\bq) & 0
                \end{pmatrix}\notag \\
                &+ \frac{\sqrt{6}}{12}t_5 \tau_0 \otimes 
                \begin{pmatrix}
                        h_2 (\bq) &h_1 (\bq) &0 \\
                        h_1 (\bq)&- h_2 (\bq) &0 \\
                        0&0 &0
                \end{pmatrix}\notag \\
                &+ \frac{\sqrt{6}}{12} i t_6 \tau_z \otimes 
                \begin{pmatrix}
                        0&0 & h_3 (\bq)\\
                        0&0&h_4 (\bq)  \\
                           -h_3 (\bq)& -h_4 (\bq)& 0
                \end{pmatrix},\notag \\
                \end{align}  
for the terms allowed in the non-ferroaxial system and 
                \begin{equation}
                H_\bk^2 = 
                \frac{\sqrt{6}}{12} t_\text{ax} \, \tau_0 \otimes 
                \begin{pmatrix}
                        -h_1 (\bq) &h_2 (\bq) & 0\\ 
                        h_2 (\bq)&h_1 (\bq)& 0\\
                        0&0&0
                \end{pmatrix},
                \end{equation}
for the term permitted in the ferroaxial phase.
Here we introduced
\begin{align}
h_1 (\bq) &= \sqrt{3}\left\{ -\cos{q_y} + \cos{(q_x+q_y)}\right\},\\
h_2 (\bq) &= 2 \cos{q_x}-\cos{q_y}-\cos{(q_x+q_y)},\\
h_3 (\bq) & = -  \left\{ 2 \sin{q_x} - \sin{q_y} + \sin{(q_x+q_y)} \right\} ,\\
h_4 (\bq) & = -\sqrt{3}  (\sin{q_y} + \sin{(q_x+q_y)}),
\end{align}
with momentum $\bq$ in the primitive lattice representation ($q_x = k_x$, $q_y = -\frac{1}{2}k_x + \frac{\sqrt{3}}{2}k_y$, and $q_z = k_z$).
More importantly, the sign of $t_\text{ax}$ represents the polarity of the ferroaxial order ($A_+$/$A_-$).


We compute the nonlinear susceptibilities $\chi_1$ and $\chi_2$.
The photoelectric field was introduced by a shift in the crystal momentum $\bk \to \bk -\bA$, where $\bE(\omega) = i \omega \bA$.
The coupling of the phonon excitation $\PH$ to electrons is desctibed by the perturbative Hamiltonian 
                \begin{equation}
                \delta H_\PH = \sum_\bk \hat{X}_\PH (\bk) \PH,
                \end{equation}
where the operator $\hat{X}_\PH$ is given phenomenologically in a symmetry-adapted form consistent with $\PH$.
The operator $\hat{X}_{\PH_i}$ is coupled to $\PH_i$, which belongs to the $\nEg{i}$-irreducible representation.
The explicit form is obtained as 
                \begin{align}
                        \hat{X}_{\PH_1} (\bk) &=  \bc_\bk^\dagger \tau_0 \otimes 
                        \begin{pmatrix}
                                 1& -i&0 \\
                                 -i& -1&0 \\
                                 0&0 &0 
                        \end{pmatrix} \bc_\bk,\\
                        \hat{X}_{\PH_2} (\bk) &=  \bc_\bk^\dagger \tau_0 \otimes 
                        \begin{pmatrix}
                                 1& i&0 \\
                                 i& -1&0 \\
                                 0&0 &0 
                        \end{pmatrix} \bc_\bk.
                \end{align}
Following the established second-order perturbation calculations, the nonlinear susceptibility is given by 
                \begin{align}
                \chi_{ij}^\PH (\omega,\delta  \omega)
                =\frac{- 1}{2\omega  (\omega +  \delta \omega)} \frac{1}{N}\sum_\bk 
                        &\Biggl[ 
                        \sum_{a,b}  \frac{J^{ij}_{ab} X^\PH_{ba} f_{ab} }{ \delta \omega +i\eta_X -\epsilon_{ba}}\cr
                        &+\sum_{a,b,c} \frac{J^{i}_{ab}}{ \omega + \delta \omega + i(\eta_J+\eta_X) -\epsilon_{ba}} \left(   \frac{J^j_{bc}X^\PH_{ca}f_{ac} }{\delta  \omega + i \eta_X -\epsilon_{ca} }    -    \frac{J^j_{ca}X^\PH_{bc}f_{cb} }{\delta  \omega + i \eta_X -\epsilon_{bc} }  \right)\cr
                        &+\sum_{a,b,c}\frac{J^{i}_{ab}}{ \omega + \delta \omega + i(\eta_J + \eta_X) -\epsilon_{ba}} \left(  \frac{J^k_{bc}X^\PH_{ca}f_{ac} }{\omega + i \eta_J -\epsilon_{ca} }    -    \frac{J^k_{ca}X^\PH_{bc}f_{cb} }{\omega + i \eta_J -\epsilon_{bc} } \right)  \Biggr],
                        \cr
                \label{nonlinear_susceptibility}
                \end{align}
where $\hat{J}^i = \partial_{k_i} \left[ H_\bk^1 + H_\bk^2 \right]$ and $\hat{J}^{ij} = \partial_{k_i} \partial_{k_j} \left[ H_\bk^1 + H_\bk^2 \right]$ are paramagnetic and diamagnetic current operators, respectively.
$N$ is the number of grid points in the discretized Brillouin zone, and $(a,b,c)$ are the band indices.
The quantities $\epsilon_{ab} = \epsilon_{\bk a} - \epsilon_{\bk b}$ and $f_{ab} = f  (\epsilon_{\bk a}) - f (\epsilon_{\bk b})$ represent the difference in energies $\epsilon_{\bk a}$ and the difference in Fermi-Dirac distribution factors between bands $a$ and $b$, respectively.
The Fermi-Dirac distribution function was parameterized by the temperature $T$ and chemical potential $\mu$.
Scattering effects are phenomenologically taken into account using the smearing factors $\eta_J, \eta_X$ for the $\bJ$ and $X$ vertices, respectively.
We use the parameters 
\begin{equation}
    t_1 = 0.5,~ t_2 = 1.0,~ t_3 = 0.3,~ t_4 = 0.3,~ t_5 = 0.2,~ t_6 = 0.1,
\end{equation}
for the non-ferroaxial hopping terms (ferroaxial parameter $t_\text{ax}$ is mentioned in each plot) and
\begin{equation}
    N = 150^3,~ \eta_J = 0.01 ,~ \eta_X = 0.01,~ T = 0.001,~\mu = -1.05.
\end{equation}

In the main text, we plot the $\omega$-dependence of the normalized cross-circular Raman scattering intensities, $\overline{I}_\text{LR}$ and $\overline{I}_\text{RL}$ [Eqs.~\eqref{normalized_LR_intensity} and \eqref{normalized_RL_intensity}], where $\delta \omega =\pm 0.001$ and $t_\text{ax} =0.1$. 
To investigate the effect of resonant particle-hole excitations, we also calculate the spectrum of the joint density of states, which is defined by
\begin{equation}
    \text{JDOS} \, (\omega) = \frac{1}{N} \sum_\bk \sum_{a,b}\frac{\pi/\gamma}{\gamma^2 + (\omega - \epsilon_{ba})^2}f_{ab}.
\end{equation}
$\gamma$ is the broadening factor concerning the resonant particle-hole creations for the excitation frequency $\omega = \epsilon_{ba}$.
In the main plot, we used $\gamma=0.01$ and $N=150^3$.
It is shown that the cross-circular ROA is strongly enhanced by the particle-hole excitations (see the comparison of cross-circular Raman scattering intensities and JDOS spectra in Fig.~4 of the main text).
The enhancement is attributed to the resonance factor such as $(\omega +  i\eta )^{-1}$ of Eq.~\eqref{nonlinear_susceptibility} in as in the case of antisymmetric Raman scatterings~\cite{Koningstein1968-ci,Mortensen1968-gn}.
We further examined the parameter dependence to verify its consistency with the experimental observations.
In Fig.~\ref{fig:model_domain_dependence}, we plot the normalized cross-circular Raman scattering intensities for different ferroaxial states ($t_\text{ax} = \pm 0.1$).
The microscopic calculations satisfy the following relationships:
\begin{equation}
    \overline{I}_\text{LR} (A_+) = \overline{I}_\text{RL} (A_-),~\overline{I}_\text{RL} (A_+) = \overline{I}_\text{LR} (A_-).
\end{equation}
Thus, our microscopic calculations reproduced the ferroaxial domain dependence of cross-circular ROA, as observed experimentally.
\begin{figure}[H]
  \begin{tabular}{cc}
  \includegraphics[width=0.5\hsize]{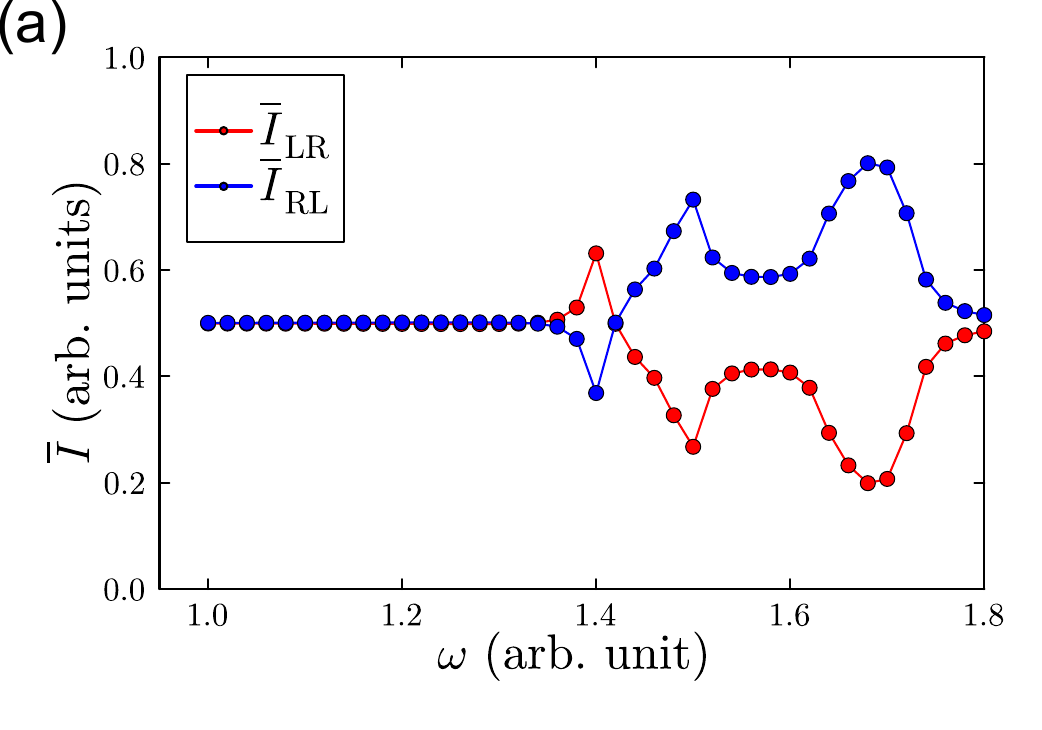}
  &
  \includegraphics[width=0.5\hsize]{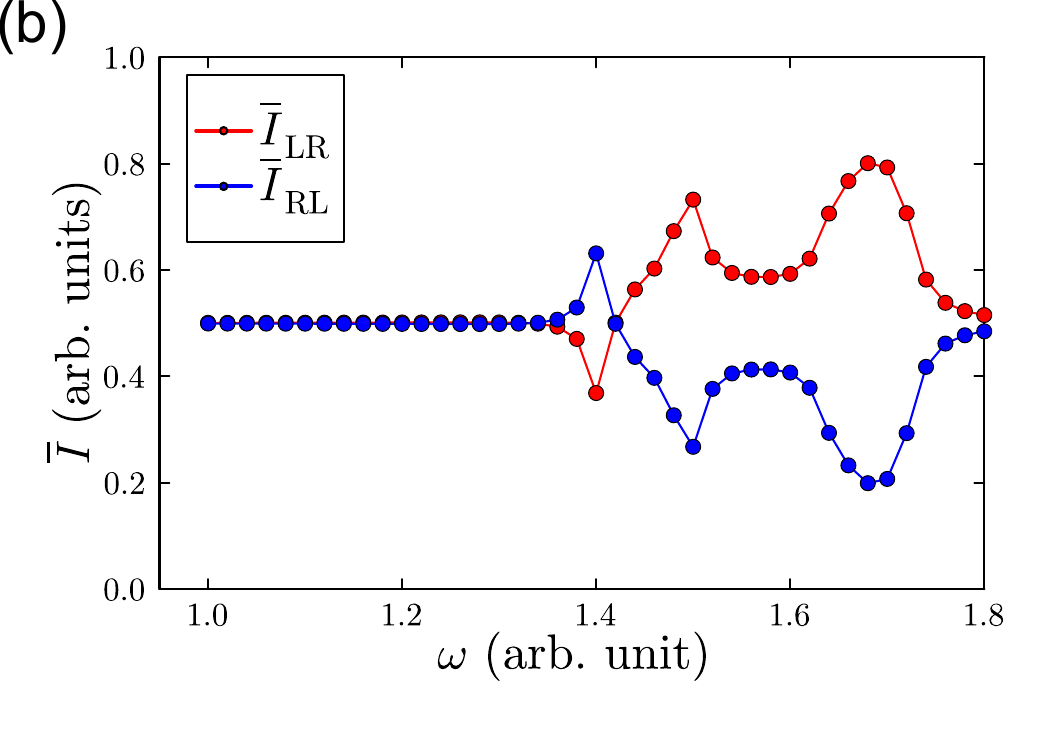}
  \end{tabular}
  \caption{\label{fig:model_domain_dependence}
  Model calculations of the excitation frequency ($\omega$) dependence of the normalized cross-circular Raman scattering intensities ($\overline{I}_\text{LR},\overline{I}_\text{RL}$) with $\delta \omega = 10^{-3}$.
  (a) Case of the ferroaxial domain state with $t_\text{ax}=0.1$,
  (b) Case of the opposite domain state with $t_\text{ax}=-0.1$.
  }
\end{figure}


\begin{figure}[H]
  \centering
  \includegraphics[width=0.98\hsize]{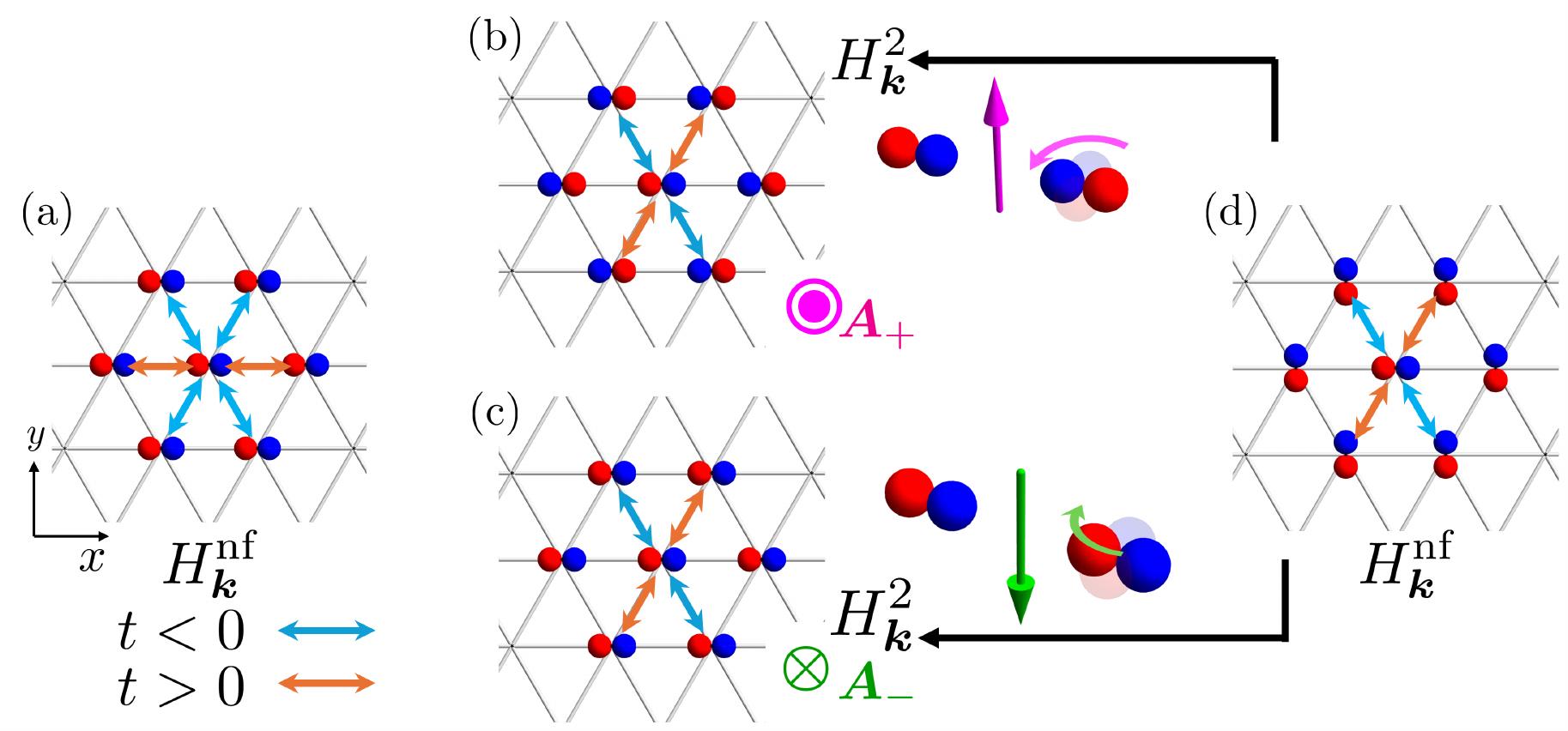}
  \caption{\label{fig:orbital_rotation}
  Comparison of (a,d) non-ferroaxial ($H_\bk^\text{nf}$) and (b,c) ferroaxial ($H_\bk^2$) hopping processes and illustration of orbital rotation driven by the ferroaxial order.
  (a) non-ferroaxial hopping process for $p_x \to p_x$. Hoppings are denoted by double-headed arrows colored in cyan ($t<0$) and orange ($t>0$).
  (b) ferroaxial hopping process for $p_x \to - p_x$ in the case of the $+z$ ferroaxial order ($A_+$). 
  (c) ferroaxial hopping process for $p_x \to p_x$ in the case of the $-z$ ferroaxial order ($A_-$). 
  (d) non-ferroaxial hopping process for $p_x \to p_y$. 
  The hopping configurations illustrated in (b,c) can be obtained by applying the orbital rotations to the $p_y$ orbitals in (d) as $p_y \to \pm p_x$.
  Note that the opposite phases of surrounding $p_x$ orbitals in (b,c) indicate the opposite ferroaxial hopping $t_\text{ax}$.  
  As a result of the staggered orbital rotations associated with opposite ferroaxial polarizations, the signs of the ferroaxial hopping parameters in (b,c) are reversed.
  }
\end{figure}

To gain insight into the ferroaxial state, we considered the hopping processes of the $p_x$ and $p_y$ orbitals on the same sublattice governed by $t_5$ and $t_\text{ax}$.
Letting the non-ferroaxial hopping described by $t_5$ be $H_\bk^\text{nf}$, the nonzero matrix elements are obtained as follows:
\begin{align}
    \Braket{p_x | H_\bk^\text{nf} | p_x} = h_2 (\bq),~
    \Braket{p_x | H_\bk^\text{nf} | p_y} = h_1 (\bq),~
    \Braket{p_y | H_\bk^\text{nf} | p_x} = h_1 (\bq),~
    \Braket{p_y | H_\bk^\text{nf} | p_y} = -h_2 (\bq),
\end{align}
where we omit the prefactors on the right-hand sides, as well as parameters of the bra and ket states, such as $\bk$ and sublattice degrees of freedom.
Similarly, the ferroaxial component $H_\bk^2$ can be explicitly written as
\begin{align}
    \Braket{p_x | H_\bk^2 | p_x} = -h_1 (\bq),~
    \Braket{p_x | H_\bk^2| p_y} = h_2 (\bq),~
    \Braket{p_y | H_\bk^2 | p_x} = h_2 (\bq),~
    \Braket{p_y | H_\bk^2 | p_y} = h_1 (\bq),
\end{align}
again omitting prefactors on the right-hand sides.
One can notice that the ferroaxial terms have forms similar to those of the non-ferroaxial terms.
Taking a ferroaxial hopping term, we apply a $-\pi/2$ rotation along the $z$-axis to the bra's orbital and replace $H_\bk^2$ with $H_\bk^\text{nf}$.
Consequently, we can reproduce the non-ferroaxial hoppings (Fig.~\ref{fig:orbital_rotation}). 
For example, in the case of $\Braket{p_x | H_\bk^2 | p_x}$,
\begin{equation}
     -h_1 (\bq) = \Braket{p_x | H_\bk^2 | p_x} \to - \Braket{p_y | H_\bk^\text{nf} | p_x} = - h_1 (\bq).
\end{equation}
This correspondence indicates that the $z$-polarized ferroaxial order gives rise to hybridization of $p_x$ and $p_y$ orbitals, which are in a transversal relation in the plane normal to the ferroaxial vector~\cite{hayami2022}.
Thus, the hopping energy $\Braket{p_y | H_\bk | p_x}$ consists of non-ferroaxial contributions such as $\Braket{p_y | H_\bk^\text{nf} | p_x} = h_1 (\bq)$, as well as the ferroaxial contribution $\Braket{p_y | H_\bk^2 | p_x} = h_2 (\bq)  \propto \Braket{p_x | H_\bk^\text{nf} | p_x} $, arising from the ferroaxial orbital rotation of the bra state, expressed by $\bra{p_y} \to \bra{p_y} + \delta \bra{p_x}$ ($\delta \propto t_\text{ax}$).


Next we analytically demonstrate that the cross-circular ROA originates from the coupling between the $E_{g}$ phonons and orbital rotations driven by the ferroaxial order, through an analysis of the essential model parameters.
To this end, we adopt the systematic framework developed in Ref.~\cite{oiwa2022a}, which enables the identification of the essential model parameters in the nonlinear susceptibility given by Eq.~(\ref{nonlinear_susceptibility}). 
The absence of these parameters leads to the disappearance of the cross-circular ROA.
By following the method introduced in Ref.~\cite{oiwa2022a}, Eq.~(\ref{nonlinear_susceptibility}) can be rewritten as
\begin{align}
    \chi_{ij}^\PH (\omega,\delta  \omega) = \sum_{l,m} \Lambda^{\omega, \delta \omega}(l,m) \Gamma_{ij}^{\PH}(l,m) + \sum_{l,m,n} \Lambda^{\omega, \delta \omega}(l,m,n) \Gamma_{ij}^{\PH}(l,m,n),
    \label{nonlinear_susceptibility_2}
\end{align}
where the first and second terms correspond to the diamagnetic term and the remaining terms in Eq.(\ref{nonlinear_susceptibility}), respectively.
$l$, $m$, and $n$ represent the integer indices.
$\chi_{ij}^\PH (\omega,\delta  \omega)$ is decomposed into two parts: $\Lambda^{\omega, \delta \omega}$ and $\Gamma_{ij}^{\PH}$. 
The latter part $\Gamma_{ij}^{\PH}$ depends only on the operators $\hat{J}^{i}$, $\hat{J}^{j}$, and $\hat{X}^{\PH}$.
$\Gamma_{ij}^{\PH}(l,m)$ and $\Gamma_{ij}^{\PH}(l,m,n)$ are given explicitly as follows:
\begin{align}
\Gamma_{ij}^{\PH}(l,m) &= \sum_{\bm{k}} \mathrm{Tr} \left(\hat{J}^{ij} H_{\bk}^{l} \hat{X}^{\pm} H_{\bk}^{m}\right), 
\label{Gamma_lm} \\
\Gamma_{ij}^{\PH}(l,m,n) &= \sum_{\bm{k}} \mathrm{Tr} \left(\hat{J}^{i} H_{\bk}^{l} \hat{J}^{j} H_{\bk}^{m} \hat{X}^{\PH} H_{\bk}^{n}\right),
\label{Gamma_lmn} 
\end{align}
where $H_{\bk}^{l}$ represents the $l$th power of $H_{\bk}$.
Thus, all the model parameters and symmetry information of $\chi_{ij}^\PH (\omega,\delta  \omega)$ are embedded in $\Gamma_{ij}^{\PH}$.
In contrast, the former part, $\Lambda^{\omega, \delta \omega}$, depends on external parameters such as the frequency, temperature, and chemical potential dependencies through $f(\epsilon)$, and the smearing factor $\eta$. 
By analyzing the latter part, we can identify the essential parameters in $\chi_{ij}^\PH (\omega,\delta  \omega)$. 
In the following, we only focus on the $\Gamma_{ij}^{\PH}$ term in order to extract the essential parameters.

We focus on the components $\chi_{i}=\chi_{xx}^{\PH_i} (\omega,\delta  \omega)$ $(i=1,\pad 2)$, which contribute to the cross-circular ROA.
By using Eq.~(\ref{nonlinear_susceptibility_2}), $|\chi_{xx}^{\PH_{1}}|^{2} - |\chi_{xx}^{\PH_{2}}|^{2}$ in Eq.(\ref{cross_circular_ROA}) is expressed as
\begin{align}
    |\chi_{xx}^{\PH_{1}}|^{2} - |\chi_{xx}^{\PH_{2}}|^{2} 
    =& -2 \sum_{l\neq l',m\neq m'} 
    \mathrm{Im} \left[\Lambda^{\omega, \delta \omega}(l,m) \Lambda^{\omega, \delta \omega *}(l',m') \right] 
    \mathrm{Im} \left[\Gamma_{xx}^{\PH_{1}}(l,m) \Gamma_{xx}^{\PH_{2}}(l',m') \right] 
    \cr &
    - 2 \sum_{l\neq l', m\neq m', n\neq n'} 
    \mathrm{Im} \left[\Lambda^{\omega, \delta \omega}(l,m,n) \Lambda^{\omega, \delta \omega *}(l',m',n') \right] 
    \mathrm{Im} \left[\Gamma_{xx}^{\PH_{1}}(l,m,n) \Gamma_{xx}^{\PH_{2}}(l',m',n') \right] 
    \cr &
    - 4 \sum_{l,m,l',m',n'} 
    \mathrm{Im} \left[\Lambda^{\omega, \delta \omega}(l,m) \Lambda^{\omega, \delta \omega *}(l',m',n') \right] 
    \mathrm{Im} \left[\Gamma_{xx}^{\PH_{1}}(l,m) \Gamma_{xx}^{\PH_{2}}(l',m',n') \right].
    \cr
    \label{ccROA}
\end{align}
Note that the real part of the product of $\Gamma_{xx}^{\PH_{1}}$ and $\Gamma_{xx}^{\PH_{2}}$ vanishes owing to spatial inversion and time-reversal symmetries.
The first, second, and third terms in Eq.~(\ref{ccROA}) are nonzero only when the following conditions are satisfied:
\begin{align}
    &\mathrm{Im} \left[\Gamma_{xx}^{\PH_{1}}(l,m) \Gamma_{xx}^{\PH_{2}}(l',m') \right] \neq 0,
    \label{Gamma_Gamma_1}
    \\
    &\mathrm{Im} \left[\Gamma_{xx}^{\PH_{1}}(l,m,n) \Gamma_{xx}^{\PH_{2}}(l',m',n') \right] \neq 0,
    \label{Gamma_Gamma_2}
    \\
    &\mathrm{Im} \left[\Gamma_{xx}^{\PH_{1}}(l,m) \Gamma_{xx}^{\PH_{2}}(l',m',n') \right] \neq 0.
    \label{Gamma_Gamma_3}
\end{align}
Let us now focus on the first term.
Since the essential model parameters are included in any pairs of $[(l,m), (l',m')]$, here we consider the two terms with the $[(l,m), (l',m')] = [(0,1), (0,2)]$ and $[(l,m), (l',m')] = [(0,1), (1,1)]$ as the lowest-order contributions.
These are explicitly given by
\begin{align}
\mathrm{Im} \left[\Gamma_{xx}^{\PH_{1}}(0,1) \Gamma_{xx}^{\PH_{2}}(0,2) \right]
&= \frac{243 \sqrt{6}}{2} t_{\rm ax} \left(2 \alpha_{2}^{2} t_{4}^{2} - 2 \alpha_{2}^{2} t_{6}^{2} + 144 t_{1} t_{3} t_{4}^{2} - 144 t_{1} t_{3} t_{6}^{2} 
\right. \cr & \left. \qquad\qquad\qquad
- t_{4}^{4} + 2 t_{4}^{2} t_{5}^{2} + 2 t_{4}^{2} t_{6}^{2} - 2 t_{5}^{2} t_{6}^{2} - t_{6}^{4}\right),
\\
\mathrm{Im} \left[\Gamma_{xx}^{\PH_{1}}(0,1) \Gamma_{xx}^{\PH_{2}}(1,1) \right]
&= 
243 \sqrt{6} t_{\rm ax} \left(- 2 \alpha_{2}^{2} t_{4}^{2} + 2 \alpha_{2}^{2} t_{6}^{2} + t_{4}^{4} - 2 t_{4}^{2} t_{5}^{2} - 2 t_{4}^{2} t_{6}^{2} + 2 t_{5}^{2} t_{6}^{2} + t_{6}^{4}\right).
\cr
\end{align}
In addition, the other low-order terms up to $l+m < 4$ and $l'+m' < 4$ in Eq.~(\ref{Gamma_Gamma_1}) and the lower-order contributions in Eqs.~(\ref{Gamma_Gamma_2}) and (\ref{Gamma_Gamma_3}), are proportional to $t_{\rm ax}$.
Thus, the essential model parameter in the cross-circular ROA is the ferroaxial parameter $t_{\rm ax}$, which is consistent with both the experimental observations and numerical calculations.

\clearpage
\section{\NoCaseChange{\addr{Scanning measurement of the multidomain \ce{NiTiO3} crystal}}}
\addr{The laser spot size used in Fig.~5(c) of the main text was \Sqty{3.3}{\um} (FWHM), and the scan step was \Sqty{20}{\um}, which therefore limits the effective spatial resolution of the image.
We performed high-resolution scans near domain walls, using \Sqty{785}{nm} excitation, \Sqty{1.0}{\um} spot size, and \Sqty{0.5}{\um} scan steps. 
The results as shown in Figs.~\ref{fig:domain_scan} show that a clear sign reversal of $\groa$ across the domain wall and a narrow region where $\groa\approx0$.}

\begin{figure}[H]
    \centering
    \includegraphics[width=0.98\hsize]{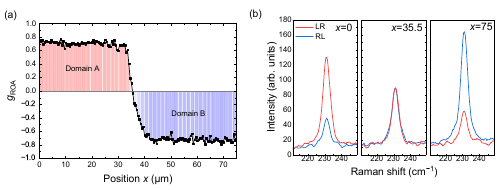}
    \caption{\label{fig:domain_scan}
    \addr{One-dimensional scanning map of $\groa$ of the $\Eg^{(1)}$ mode, and 
    (b) Raman spectra of LR and RL configurations at the position $x=\Sqty{0}{\um}$, \Sqty{35.5}{\um}, and \Sqty{75}{\um}.}}
\end{figure}

\clearpage
\section*{\NoCaseChange{\addr{Supplementary movies}}}

\addr{\textbf{Supplementary Movie 1.}
  Animation of the five $\Eg$ phonons separated into $\nEg{1}$ and $\nEg{2}$ modes.}

}

\end{document}